\documentclass[12pt,preprint]{aastex}

\begin{document}

\title{Clustered and Triggered Star Formation in W5: Observations with {\it Spitzer}}

\author{Xavier P. Koenig,\altaffilmark{1} Lori E. Allen,\altaffilmark{1} Robert A. Gutermuth,\altaffilmark{1} Joseph L. Hora,\altaffilmark{1} Christopher M. Brunt,\altaffilmark{2} James Muzerolle\altaffilmark{3}}
\altaffiltext{1}{Harvard-Smithsonian Center for Astrophysics, 60 Garden Street, 
Cambridge, MA 02138}
\altaffiltext{2}{The School of Physics, University of Exeter, The Queens Drive, Exeter, Devon, UK, EX4 4QL}
\altaffiltext{3}{Steward Observatory, 933 North Cherry Avenue, University of Arizona, Tucson, AZ 85721}

\begin{abstract}
  We present images and initial results from our extensive {\it
    Spitzer} Space Telescope imaging survey of the W5 H\,{\sc ii}
  region with the Infrared Array Camera (IRAC) and Multiband Imaging
  Photometer for {\it Spitzer} (MIPS). We detect dense clusters of
  stars, centered on the O stars: HD 18326, BD +60 586, HD 17505 and
  HD 17520. At 24 $\micron$, substantial extended emission is visible,
  presumably from heated dust grains that survive in the strongly
  ionizing environment of the H\,{\sc ii} region. With photometry of
  more than 18000 point sources, we analyze the clustering properties
  of objects classified as young stars by their IR spectral energy
  distributions (a total of 2064 sources) across the region using a
  minimal-spanning-tree algorithm. We find $\sim$40--70\% of infrared
  excess sources belong to clusters with $\geq$10 members. We find
  that within the evacuated cavities of the H\,{\sc ii} regions that
  make up W5, the ratio of Class II to Class I sources is $\sim$7
  times higher than for objects coincident with molecular gas as
  traced by $^{12}$CO emission and near-IR extinction maps. We
  attribute this contrast to an age difference between the two
  locations, and postulate that at least two distinct generations of
  star formation are visible across W5. Our preliminary analysis shows
  that triggering is a plausible mechanism to explain the multiple
  generations of star formation in W5, and merits further
  investigation.
\end{abstract}


\section{Introduction}
Star formation is a self-regulating process---once massive stars form
they immediately begin to disrupt their natal environment with their
stellar winds and the emission of ionizing radiation. Eventually their
parental molecular clouds are destroyed, halting further star
formation. However, it has also been argued that the energy input by
these massive stars can promote and induce subsequent star formation
in the surrounding molecular gas before it disperses, above that which
would be produced without external forcing. This process is given the
name `triggering' \citep[see][for reviews of this
subject]{elmegreen98,zinnecker07}. It is vital to understand the
balance of cloud destruction and triggered star formation if we are to
develop a theory that explains the morphology and evolution of star
forming regions, star formation efficiencies and the initial mass
function of stars in clusters. It also has relevance for star
formation on galactic scales in understanding how star formation
progresses with time, and indeed how spiral structure of galaxies
evolves with time \citep{seiden90,jungwiert94}. Feedback in star
formation can be clearly observed in bright-rimmed clouds, where an
edge-on molecular cloud is externally illuminated by nearby young
massive stars, creating a cross-section of the photo-evaporation
process as the ionization fronts they produce advance into the
molecular cloud.

Two triggering mechanisms are of interest in regard to W5. The first
is the creation of an ionized H\,{\sc ii} region bubble by an initial
generation of massive stars within their parental molecular cloud, and
its subsequent expansion. In this `collect and collapse' mechanism,
investigated analytically by \citet{whitworth94} and numerically by
\citet{dale07}, the expansion creates a shock front that sweeps up
neutral material ahead of it as it progresses outward. The gas
accumulated eventually exceeds a critical threshold for collapse and
gives rise to a second generation of star formation. Secondly,
inhomogeneity in the ISM often leads to small clumps of material
remaining exposed inside an H\,{\sc ii} region. In this environment,
the high pressure of the surrounding ionized gas has been suggested as
a mechanism to compress the clumps and form stars \citep{stutzki88}.

The star forming region W5 is a part of the chain of molecular clouds
W3/4/5 \citep{westerhout58}. Its distance has not been definitively
established---\citet{becker71} found a photometric distance of 2.2
kpc, but more recently, \citet{hillwig06} found that a distance of 1.9
kpc gave the most consistent results of evolutionary model fits to
stellar radii in W5. \citet{xu06} found a distance to the neighboring
region W3OH of 1.95$\pm$0.04 kpc using maser parallaxes---we adopt a
distance to W5 of 2 kpc as a conservative intermediate between these
different estimates. W5 is a relatively isolated star forming region,
with an apparently simple morphology. Optical imaging shows it is made
up of two roughly circular adjoining H\,{\sc ii} regions W5 East and
W5 West \citep{karr03}, containing one and four O stars
respectively. At least two of the four O stars in W5 West are multiple
systems \citep{hillwig06}. Both $^{12}$CO and 21 cm radio emission
\citep[see][]{normandeau96} demonstrate the same overall
shape---$^{12}$CO emission traces the molecular hydrogen gas in W5,
which appears as two broken rings of emission. Figure 1 shows a map of
$^{12}$CO ($\lambda$ = 2.6 mm) emission integrated over the range
$-$28 to $-$47 km s$^{-1}$, made using observations with the 14m FCRAO
telescope. The bulk of the emission is found between a velocity of
$-$31 and $-$47 km s$^{-1}$, and thus is likely roughly in the same
plane. In this image, following the naming scheme of
\citet{wilking84}, to the northwest of HD 17505 is the molecular cloud
W5NW, in between HD 18326 and BD +60 586 is W5NE, and to the east of
HD 18326 is W5A. Due to this relatively simple morphology, W5 presents
a useful test case for investigating models of triggered star
formation and the influence of massive star formation on its
surroundings.

\begin{figure}
\begin{center}
\includegraphics[width=4.5in]{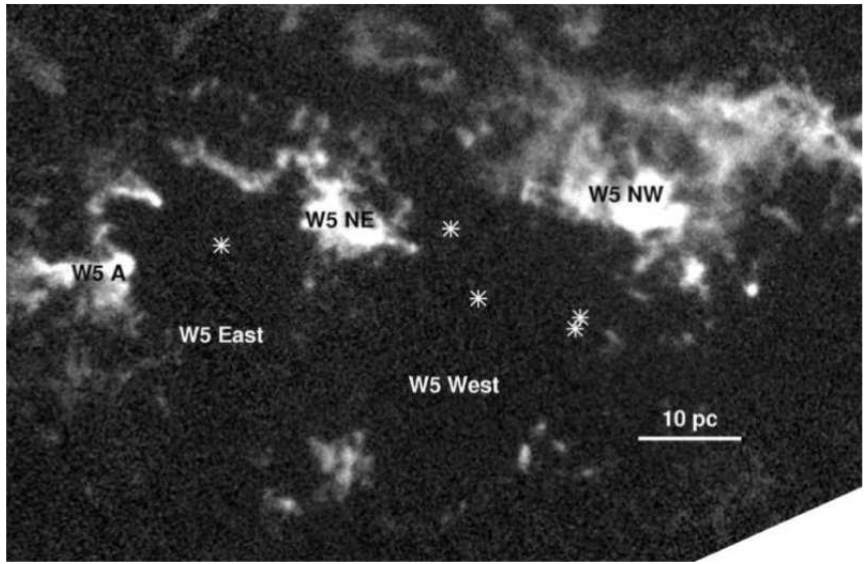}
\caption{$^{12}$CO (1-0) greyscale image of W5 from FCRAO telescope
  (integrated T$_A^*$ corrected for main beam efficiency
  $\eta_{MB}=0.45$, ranging from 0--110 K km s$^{-1}$), oriented
  North-up and East-left. We label molecular clouds in black following
  the naming scheme of \citet{wilking84}. We label the two H\,{\sc ii}
  regions in white. Prominent O stars are marked with white
  asterisks: HD 18326 in W5 East, and in W5 West from North-East to
  South-West are BD +60 586, HD 237019 and HD 17505 (upper of pair), HD
  17520 (lower of pair).}
\end{center}
\end{figure}

In this paper we present initial results from our mid-IR {\it Spitzer}
survey of W5 with the IRAC and MIPS instruments. In $\S$ 2 we describe
our observations and data reduction techniques and our classification
and clustering analysis in $\S$ 3. In $\S$ 4 we perform a trial,
simplified investigation of triggered star formation models as a means
to explain the observed distribution of young stars in different
evolutionary states across the region. In $\S$ 5 we present our
conclusions and directions of future work.

\section{Observations and Data Reduction}
W5 was observed with the {\it Spitzer} IRAC instrument \citep{fazio04}
in all four bands (3.6, 4.5, 5.8 and 8.0 $\micron$). The observations
were broken down into three rectangular Astronomical Observing
Requests (AORs) covering $\sim$1.8$\times$1.6 degrees, in order to
observe at multiple rotation angles and help minimize artifacts
aligned along columns or rows of the array. In Table 1 we list the
dates and coordinates of each AOR. Each AOR had a coverage of 1 High
Dynamic Range (HDR) frame. HDR mode results in a 10.4 second and 0.4
second exposure being taken at each position in each map. We used
software tools ({\it clustergrinder}) developed by one of us (RAG) to
produce final image mosaics from these data in each wavelength
band. {\it Clustergrinder} incorporates all necessary image treatment
steps, for example, saturated pixel processing and distortion
corrections \citep[see][for a more complete description of the
processing performed]{gutermuth08}. {\it Clustergrinder} uses the
short 0.4 sec exposures only in saturated or near-saturated regions,
so that the combined map has an effective total integration time of
3$\times$10.4 = 31.2 sec in most of the overlapping areas. We also
incorporated in our data processing archival data covering AFGL 4029
\citep[from {\it Spitzer} GTO program PID 201][]{allen05}, at the
eastern end of W5.

\begin{deluxetable}{lllc}
\tablenum{1}
\tablewidth{0pt} 
\tabletypesize{\scriptsize}
\tablehead{ \colhead{AORKEY} & \colhead{Date} & \colhead{Coordinates} & \colhead{IRAC Reduction}\\ \colhead{ } & \colhead{(UT)} & \colhead{(J2000)} & \colhead{Pipeline ver.}}
\startdata 
\dataset[ADS/Sa.Spitzer#0014507776]{14507776} & 2006 September 20 & RA 02 53 32.9  & S14.0.0 \\
\nodata & \nodata & Decl. +60 25 59.1 & \nodata\\
\dataset[ADS/Sa.Spitzer#0014507008]{14507008} & 2006 September 28 & RA 02 53 43.5 & S14.0.0\\
\nodata & \nodata & Decl. +60 21 57.2 & \nodata\\
\dataset[ADS/Sa.Spitzer#0014508544]{14508544} & 2007 February 16 & RA 02 53 54.2  & S15.3.0 \\
\nodata & \nodata & Decl. +60 17 55.2  & \nodata
\enddata
\end{deluxetable}

The MIPS \citep{rieke04} observations were carried out on 2006
February 23 UT under our GO-2 program, PID 20300. Images were taken in
scan map mode using the medium scan speed for an average exposure time
of 41.9s per pixel once frames were combined. The raw data were
processed with pipeline version S13.2.0. We produced final mosaics
using the MIPS instrument team Data Analysis Tool, which calibrates
the data and applies a distortion correction to each individual
exposure before combining \citep{gordon05}. We used only the 24
$\micron$ band data for our analysis in this paper, since strong
background emission dominates at the longer wavelength (70 and 160
$\micron$) bands of MIPS, and lower sensitivity reduces the number of
detectable objects to an un-useful level for the present study.

In Figure 2 we present a 3-color IRAC image of W5, using the color
scheme: blue=3.6 $\micron$, green=4.5 $\micron$ and red=8.0
$\micron$. In Figure 3 we show our MIPS 24 $\micron$ mosaic. Figure 4
presents a composite image incorporating MIPS, with blue=4.5
$\micron$, green=5.8 $\micron$ and red=24 $\micron$ (MIPS).

\begin{figure}
\begin{center}
\includegraphics[angle=90,width=6.1in]{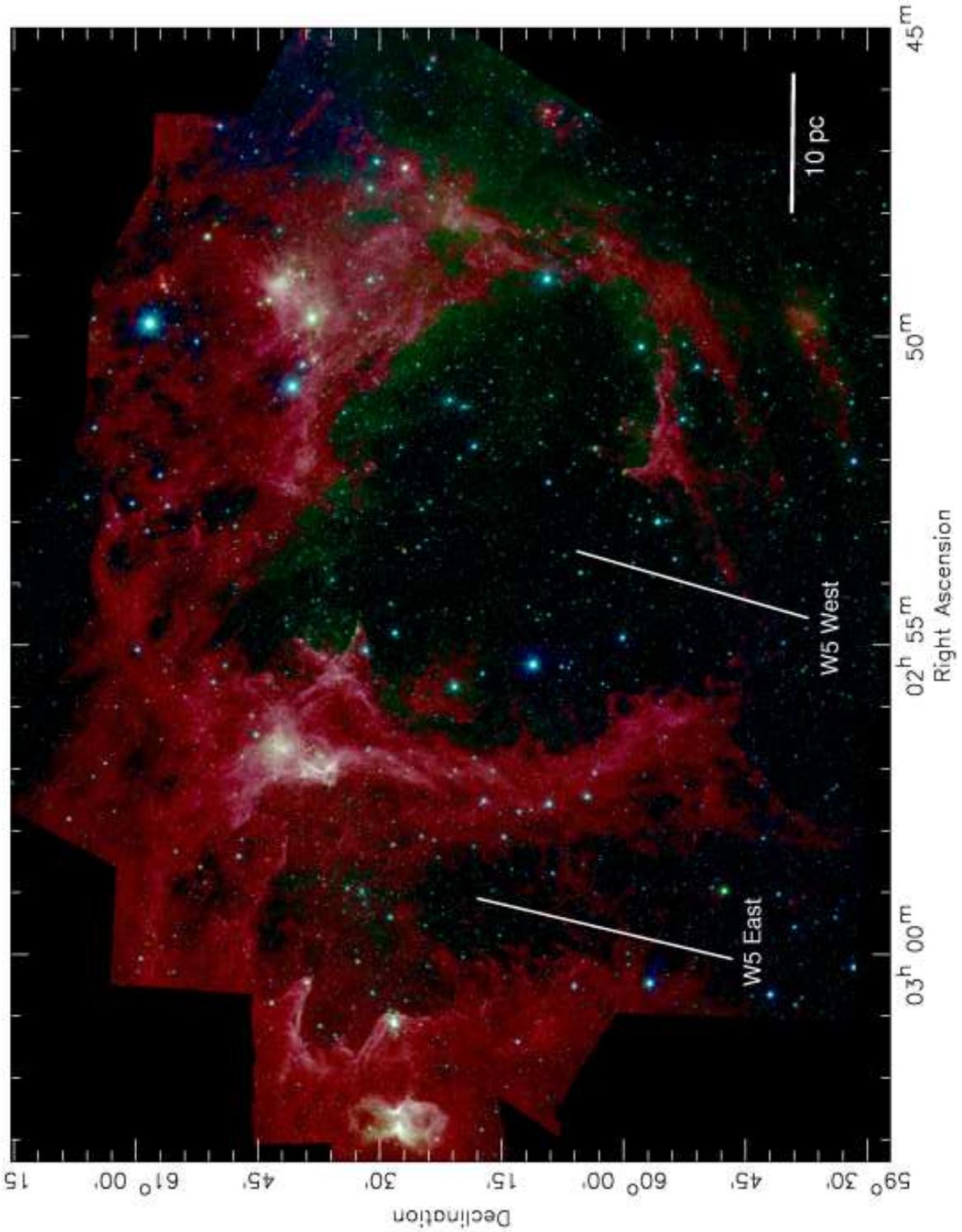}
\caption{W5 {\it Spitzer} IRAC bands 1, 2 and 4 color
  composite. Bright emission from PAH grains at 8 $\micron$ traces
  the boundary of the H\,{\sc ii} region giving the bright pinkish
  color. Dense clusters of stars surround the O stars in the interior,
  with smaller clusters appearing amidst the diffuse PAH emission.}
\end{center}
\end{figure}

\begin{figure}
\begin{center}
\includegraphics[angle=90,width=6.1in]{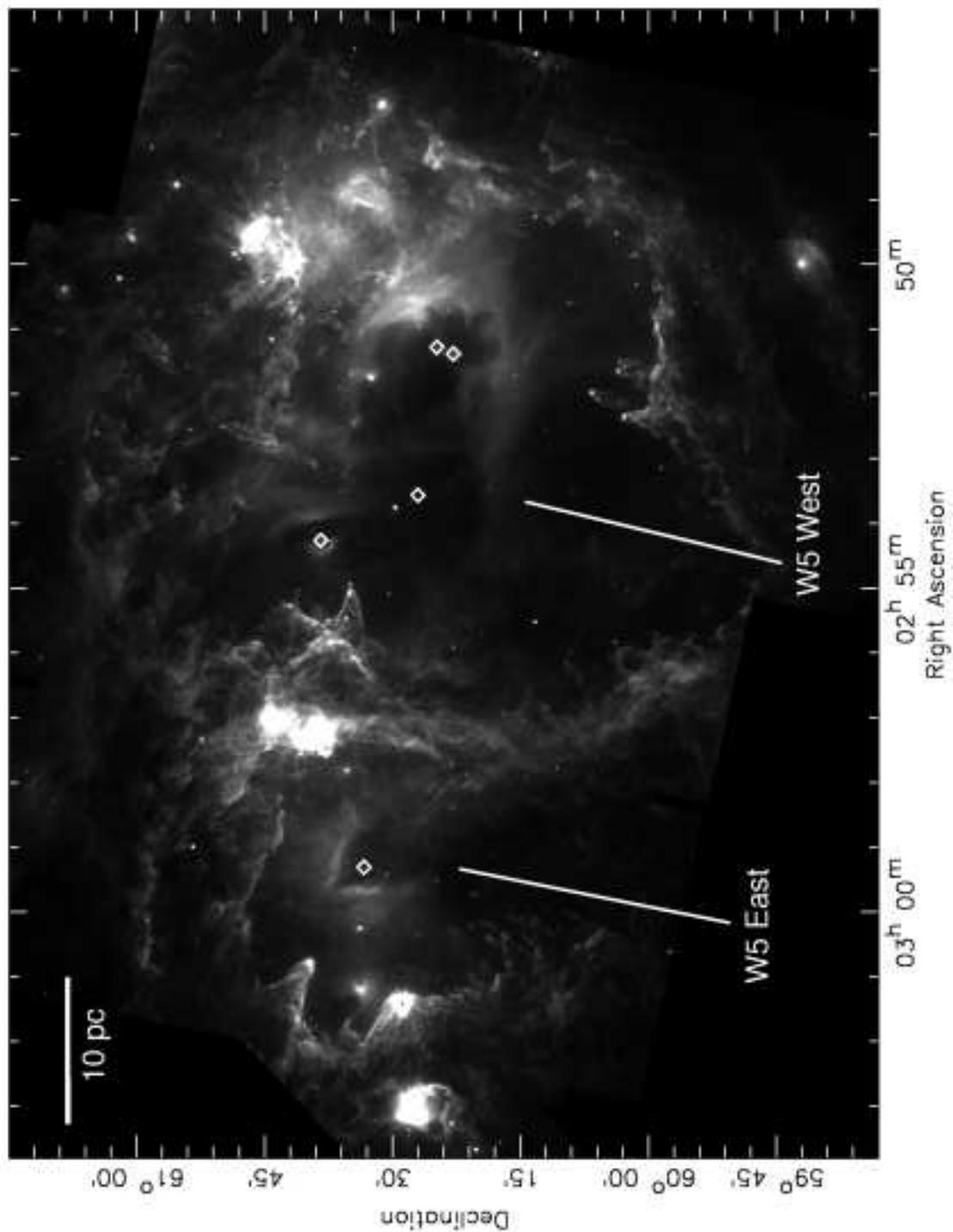}
\caption{{\it Spitzer} MIPS 24 $\micron$ mosaic. The O stars labelled
  in Fig. 1 are marked here with white diamonds. Bright dust continuum
  emission traces the boundary of the H\,{\sc ii} region, while
  extended emission from heated dust grains is also present in the
  interior (see $\S$ 3.2). Bright point sources are protostars.}
\end{center}
\end{figure}

\begin{figure}
\begin{center}
\includegraphics[angle=90,width=6.1in]{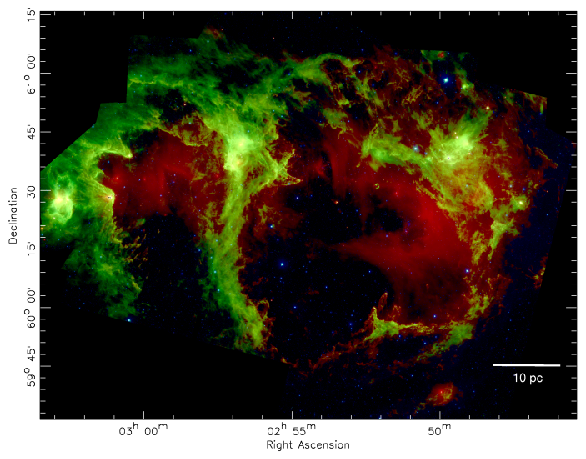}
\caption{Composite of emission in IRAC bands 2 and 4, plus MIPS 24
  $\micron$. Emission at 24 $\micron$---presumably from heated
  dust---fills the interior of the H\,{\sc ii} region cavities, see
  $\S$ 3.2.}
\end{center}
\end{figure}

We carried out point source extraction and aperture photometry of all
point sources on the final IRAC mosaics with PhotVis version
1.10beta3. PhotVis is an IDL GUI-based photometry visualization tool
\citep[see][]{gutermuth04} that utilizes DAOPHOT modules ported to IDL
in the IDL Astronomy User's Library \citep{landsman93}. We used
PhotVis to visually inspect the detected sources in IRAC bands 1, 3
and 4, adding sources not detected automatically, but clearly visible
in the images with the GUI tool and rejected any structured nebulosity
or cosmic rays mistaken for stellar sources by the automatic detection
algorithm. To save time, we did not visually check the band 2
photometry in this manner. Instead, we took the cleaned band 1 source
list as the start point for finding objects in the image and extracted
photometry at each position. Radii of the apertures and inner and
outer limits of the sky annuli were 2.4$''$, 2.4$''$ and 7.2$''$
respectively. The resultant photometry was calibrated using
large-aperture in-flight measurements of standard stars, with an
appropriate aperture correction in each channel to correct for the
smaller apertures used in this study. In this paper we restrict our
source catalog to only those objects with magnitude error $\leq$ 0.2
in all four IRAC filters, a total of 18518 objects.

We estimate the completeness in each IRAC filter by breaking up each
image into a 100 by 100 pixel grid, adding artificial stars to each
grid cell and counting the number retrieved as a function of
magnitude. Averaged over the whole W5 field, our source catalog is
90\% complete to a magnitude of 15.5 at 3.6 $\micron$, 15.5 at 4.5
$\micron$, 14.0 at 5.8 $\micron$ and 12.7 at 8 $\micron$. The
completeness is less in regions of bright diffuse emission: $\sim$14
at 3.6 and 4.5 $\micron$, $\sim$11 at 5.8 $\micron$ and $\sim$9.5 at
8.0 $\micron$.

We conducted point source extraction and aperture photometry of point
sources in the 24 $\micron$ MIPS mosaic using the
point-spread-function fitting capability in IRAF DAOPHOT
\citep{stetson87}. We visually inspected the image to pick out point
sources not automatically detected due to bright diffuse emission
evident throughout the image. We match the four-band IRAC source list
to the MIPS catalog using a 2$''$ search radius, selecting the object
closest to the MIPS point spread function centroid in cases where more
than one IRAC object is a match. Of the 1874 MIPS objects that match
with entries in our IRAC 4-band list, 6 had a second object within the
2$''$ search radius which possibly contributes to its 24 $\micron$
flux. We consider this a small effect on our analysis.

\section{Analysis}
\subsection{Source Classification}
In order to characterize the progress of star formation throughout W5
it is important to establish the evolutionary status of stars within
the region. The presence of massive O stars \citep{hillwig06} and
significant molecular material in W5 \citep{lada78}, suggest that the
stars associated with the region will be of relatively young age ($<$
10$^7$yrs). Recently formed stars exhibit infrared excess emission in
their spectra above that produced by the stellar photosphere. This
excess emission arises from heated dust, either within the
circumstellar material close to the young star, perhaps falling onto
it as a part of protostellar collapse, or left behind after the end of
star formation. This material gradually disappears with time, and
hence the excess emission evolves as a function of stellar age. We
make use of this by measuring $\alpha _{IR}$ \citep[see,
e.g.][]{lada87,stahler05}:

\begin{equation}
\alpha _{IR} = \frac{d\log \left( \lambda F_{\lambda}\right)}{d\log \lambda}
\end{equation}

the value of which decreases with the progression of the star's
evolutionary state, whether through its own aging, or the effects of
its environment.

In the case of most surveys of star forming regions, including W5,
infrared colors serve as proxies for directly measuring $\alpha
_{IR}$. Schemes developed in \citet{whitney03}, \citet{allen04},
\citet{megeath04} and \citet{muzerolle04} and tested by (for example)
\citet{hartmann05} have demonstrated the power of this technique for
classifying young stellar objects using {\it Spitzer} IRAC and MIPS
photometry. The classification scheme we use is described fully in
\citet{gutermuth08}. We use an iterative method combining IRAC and
MIPS photometry with an extinction map generated from 2MASS near-IR
photometry \citep{skrutskie06}. The scheme classifies the stars and
filters out extra-galactic contamination.

We generate an extinction map from 2MASS $H - K_S$ colors of stars
detected across the entire W5 field using the method described in
\citet{gutermuth05} which itself is based on the NICE and NICER
algorithms of \citet{lada94} and \citet{lombardi01}. The map has an
angular resolution of $\sim$35 arcsec, and is sensitive up to $A_V
\sim$15. However, the map is limited by the sensitivity of the 2MASS
survey. As a result, $A_V$ values in the map above $\sim$3 may be
underestimates of the true extinguishing column through the entire
cloud, since objects behind are too faint to be detected. This is only
the case for $<$1\% of pixels in the map, so we ignore this effect for
our analysis. With $A_V$ values in hand for all sources, we deredden
their IRAC magnitudes using the IR extinction law presented in
\citet{flaherty07}, assuming each source is seen behind the full
extinguishing column in each case.

We first remove star forming (`PAH') galaxies and weak line AGN via a
series of cuts in the four band IRAC color-color diagrams after the
procedure developed by \citet{gutermuth08}. Because W5 is
$\sim$4$\times$ more distant than the nearby regions studied by
Gutermuth et al., this filter removes many apparent young stellar
objects (YSOs). In the Appendix we discuss how we characterize and
account for this effect in our subsequent analysis. Next we filter out
unresolved shocked blobs of PAH emission by cutting objects with a
large 4.5 $\micron$ excess, i.e. very red $[3.6]-[4.5]$ color
\citep[][]{smith06}. We categorize the remaining---presumably
stellar---sample primarily relying largely on the $[4.5]-[5.8]$ color
to discriminate among SED classes. In Figure 5 we show IRAC
color-color diagrams for sources in W5, marking the location of
protostars (Class I/0, red dots), stars with disks (Class II, green
dots) and stars exhibiting only photospheric colors (Class III, black
dots) as identified using our scheme.

We also use 2MASS $H$ and $K_S$ photometry, combined with IRAC 3.6 and
4.5 $\micron$ data to classify objects lacking either an IRAC $[5.8]$
or $[8.0]$ detection. To make sure the 2MASS sources have reliable
photometry we require a magnitude error $\leqslant$0.1 in both $H$ and
$K_S$ bands. We first deredden the photometry in these four bands, and
then identify IR excess sources as those having red $[3.6]-[4.5]$ and
$K_S-[3.6]$ colors. The results are shown in Figure 6, left panel.

MIPS 24 $\micron$ photometry provides us with additional
classification information. Figure 6 (right panel) shows the
color-color diagram combining MIPS and IRAC photometry. Stellar
sources not classified as either Class I or Class II by their IRAC or
near-IR colors may still have red colors ($[5.8]-[24]$ $> $1.5) and
thus be candidate ``cold disk'' or ``transition disk'' objects---in
other words: Class II objects with significant clearing of their inner
disk region \citep[see for
example,][]{muzerolle04,lada06,cieza07,najita07,flaherty08,brown08}. These
are marked with yellow dots in Figures 5 and 6. MIPS photometry is
also used as a check on the AGN/galaxy/shocked blob filtering, and
Class I classification. AGN candidates with bright MIPS detection can
be re-classified as protostars, and Class I objects with
insufficiently red $[5.8]-[24]$ or $[4.5]-[24]$ colors are `demoted'
to Class II.

\begin{figure}
\begin{center}
\includegraphics*[width=3.0in]{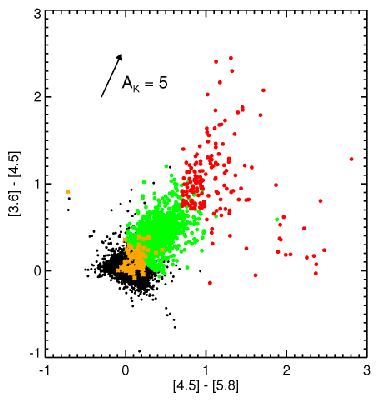}
\includegraphics*[width=3.0in]{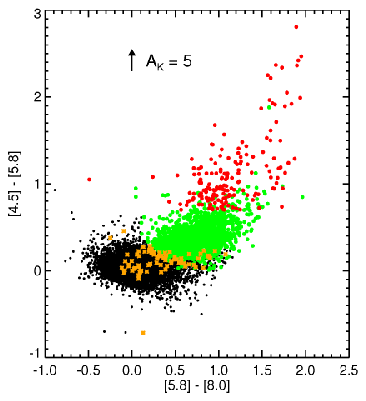}
\caption{Left panel: $[3.6]-[4.5]$ vs. $[4.5]-[5.8]$; right panel:
  $[4.5]-[5.8]$ vs. $[5.8]-[8.0]$ IRAC color-color diagrams used for
  identifying candidate protostars. Black dots---Class III,
  green---Class II, red---Class I, yellow---transition disk
  candidates. AGN and PAH galaxy candidates are not included here.}
\end{center}
\end{figure}

\begin{figure}
\begin{center}
\includegraphics*[width=3in]{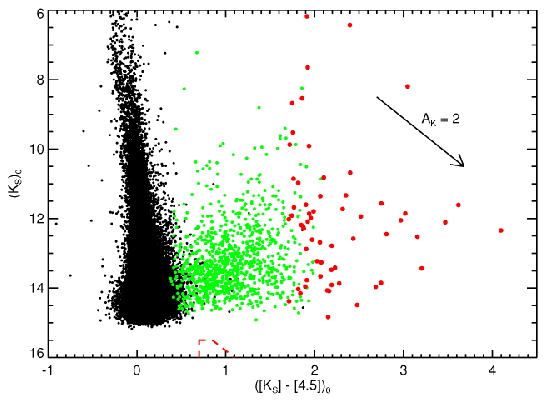}
\includegraphics*[width=3in]{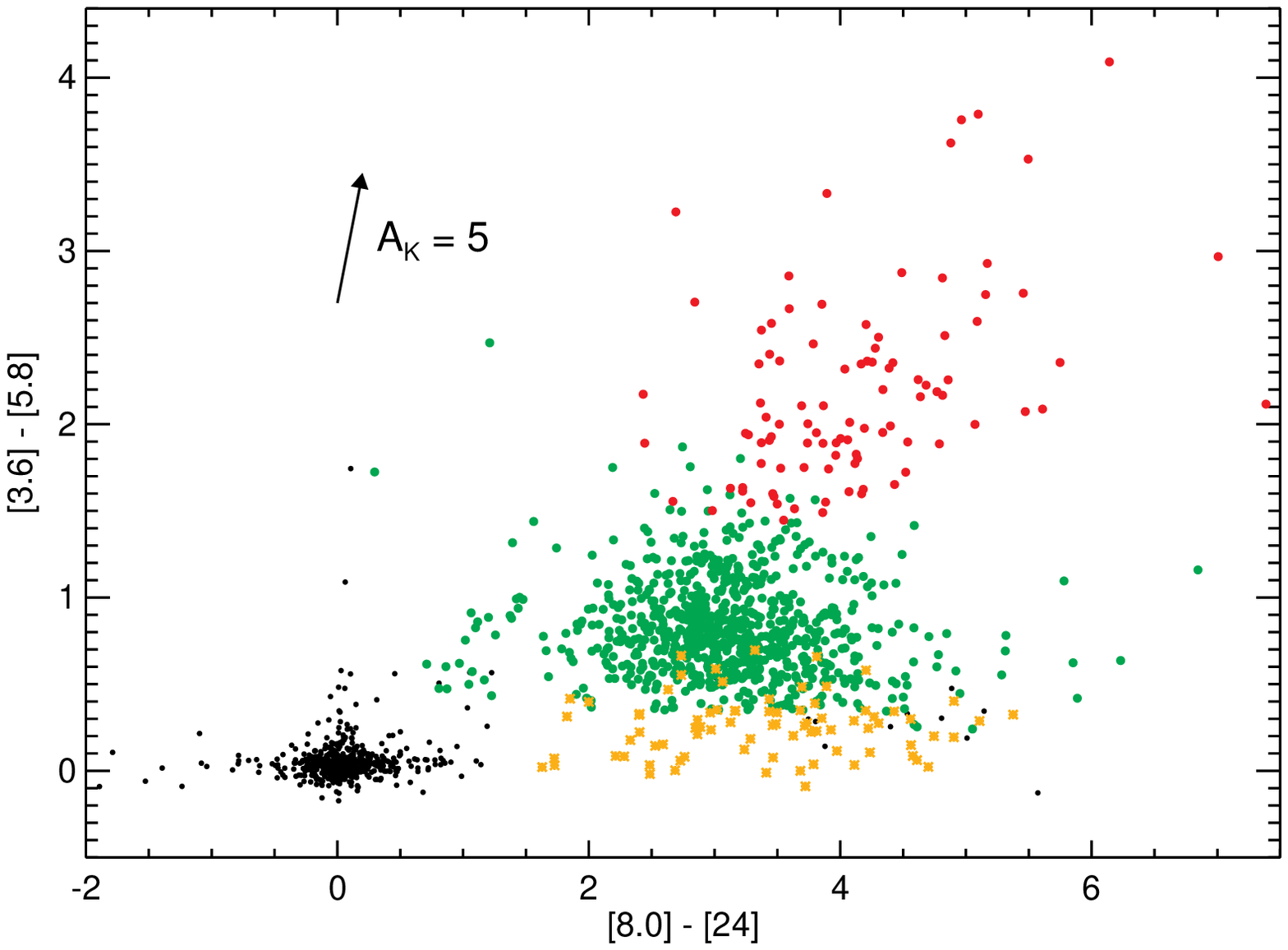}
\caption{Left panel: dereddened 2MASS $K_S$ vs. $K_S-[4.5]$
  color-magnitude diagram. Dashed red box shows location of AGN
  detected by IRAC Shallow survey \citep{eisenhardt04}. Right panel:
  IRAC and MIPS $[3.6]-[5.8]$ vs. $[8.0]-[24]$ color-color
  diagram. Color scheme: black dots---Class III, green---Class II,
  red---Class I, yellow---transition disk candidates. Suspected AGN
  and PAH galaxies are not included here.}
\end{center}
\end{figure}

Lastly, we also visually inspected the SEDs of all objects in Class I
and Class II (a total of 2064 objects), where available making use of
2MASS $JHK_S$ and/or 24 $\micron$ photometry. In most cases only IRAC
4-band data were present. We give the full list of objects: Class I,
Class II, Class III and transition disk candidates in Table 2. We list
source coordinates and photometry in near-IR $JHK_S$ (from 2MASS) and
{\it Spitzer} bands, and give a column denoting infrared SED source
class and a `flag' column for objects with unusual SEDs. Objects are
sorted and indexed by ascending right ascension order. Note that the
full machine-readable table is available in the online edition of this
paper. We note Class I sources with bright emission at 24 $\micron$
and very red IRAC$-$24 $\micron$ color ($[X]-[24] > 4.5$, where $X$ is
the magnitude in any of the four IRAC bands) as a subclass: `deeply
embedded protostars' in the table. Table 3 gives a summary of the
classification results.

\begin{deluxetable}{lcccccccccccc}
\tabletypesize{\scriptsize}
\rotate
\tablenum{2}
\tablewidth{0pt}
\tablecaption{W5 Source List}
\tablehead{\colhead{ } & \colhead{RA} & \colhead{Dec} & \colhead{$J$} & \colhead{$H$} & \colhead{$K_S$} & \colhead{$[3.6]$} & \colhead{$[4.5]$} & \colhead{$[5.8]$} & \colhead{$[8.0]$} & \colhead{$[24]$} & \colhead{ } & \colhead{ } \\
\colhead{ID} & \colhead{(deg)} & \colhead{(deg)} & \colhead{(mag)} & \colhead{(mag)} & \colhead{(mag)} & \colhead{(mag)} & \colhead{(mag)} & \colhead{(mag)} & \colhead{(mag)} & \colhead{(mag)} & \colhead{Type\tablenotemark{a}} & \colhead{Flag\tablenotemark{b}}}
\startdata 
1 & 41.081526 & 60.510607 & 15.34(06) & 13.69(03) & 13.04(03) & 12.60(01) & 12.50(01) & 12.36(04) & 12.47(07) & \nodata & III & - \\
2 & 41.098560 & 60.682772 & 10.90(02) & 10.68(02) & 10.58(02) & 10.50(01) & 10.49(01) & 10.42(01) & 9.96(02) & \nodata & III & - \\
3 & 41.100737 & 60.667424 & 13.30(02) & 12.26(02) & 11.88(02) & 11.63(01) & 11.72(01) & 11.43(04) & 10.90(11) & \nodata & III & - \\
4 & 41.109056 & 60.660875 & 13.01(02) & 12.52(04) & 12.29(03) & 12.18(01) & 12.12(01) & 11.81(05) & 10.67(05) & \nodata & II & - \\
5 & 41.109473 & 60.515686 & 12.52(02) & 12.09(02) & 11.98(02) & 11.94(01) & 11.94(01) & 11.88(02) & 12.02(05) & \nodata & III & - \\
6 & 41.112052 & 60.519115 & 13.27(02) & 12.53(02) & 12.12(02) & 11.95(01) & 11.86(01) & 11.75(02) & 11.74(05) & \nodata & III & - \\
7 & 41.116165 & 60.509403 & 14.20(03) & 13.73(03) & 13.53(04) & 13.48(01) & 13.52(01) & 13.54(08) & 13.43(17) & \nodata & III & - \\
8 & 41.128907 & 60.512183 & 14.90(04) & 14.12(04) & 13.67(05) & 13.41(01) & 13.32(01) & 13.36(08) & 13.78(19) & \nodata & III & - \\
9 & 41.128946 & 60.536046 & 13.96(03) & 12.39(02) & 11.74(02) & 11.36(01) & 11.35(01) & 11.20(01) & 11.12(03) & \nodata & III & - \\
10 & 41.130516 & 60.673029 & 12.78(03) & 12.30(03) & 12.03(03) & 11.89(01) & 11.82(01) & 11.74(02) & 11.70(05) & \nodata & III & - \\
\enddata
\tablecomments{Table 2 is published in its entirety in the electronic edition of the {\it Astrophysical Journal}. A portion is shown here for guidance regarding its form and content. Values in parentheses by photometry signify error in last 2 digits of magnitude value. Right ascension and Declination coordinates are J2000.0.}
\tablenotetext{a}{Source class as defined in text.}
\tablenotetext{b}{Sources with unusual SED shape flagged with `v' or uncertain source class `?'}
\end{deluxetable}

\begin{deluxetable}{lr}
\tablenum{3}
\tablewidth{0pt} 
\tabletypesize{\scriptsize}
\tablecaption{Source classification summary}
\tablehead{ \colhead{Class} & \colhead{N Objects}}
\startdata 
Class I/Protostars & 176\\ 
Class II & 1806 \\ 
Class III/Photospheres & 15709\\ 
Trans. Disk & 79\\ 
Embedded Protostars & 3\\
AGN & 729\\
PAH Galaxies & 198\\ 
\hline Total & 18518\enddata
\end{deluxetable}

In Figure 7 we show the spatial distribution of young stars (Class I
and deeply embedded protostars, and Class II and transition disk
candidates) in W5 overlaid on the IRAC 4.5 $\micron$ image. Class I
and Class II sources are marked in red and green respectively,
transition disk candidates and embedded protostars with yellow and
blue dots respectively. O stars are labelled and marked with white
asterisks.

\begin{figure}
\begin{center}
\includegraphics[angle=90,width=6.1in]{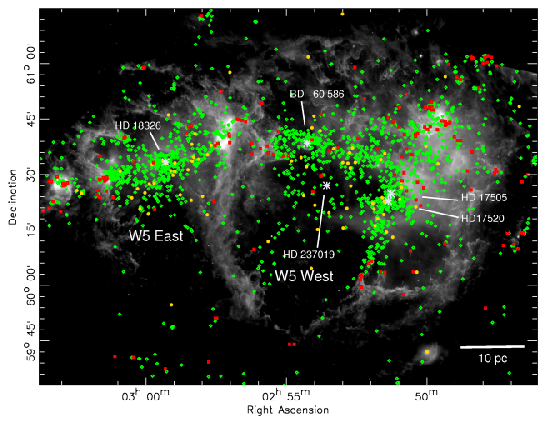}
\caption{Source distribution overlaid on the MIPS 24$\micron$
  greyscale image. Color scheme: red---Class I, green---Class II,
  yellow dots---transition disk candidates. O stars marked with
  asterisks.}
\end{center}
\end{figure}

\subsection{Spatial Distribution of Young Stars in W5}
Figure 7 clearly shows that neither Class I nor Class II type objects
in W5 are distributed uniformly, but rather (for the most part) in
clustered, or filamentary formations. Within the larger cavity of W5W,
clusters of Class II YSOs can be seen centered on the O stars HD
17505, HD 17520 and BD +60 586 of the open cluster OCl 364
\citep{alter70}. Similarly in the smaller W5E bubble, a cluster of
young stars is visible centered on the O star HD 18326. Smaller
clusters of Class I and Class II objects are also visible in the large
region of bright diffuse emission to the North-West of HD17505. Based
on the higher proportion of Class I stars relative to Class II's
(quantified in $\S$3.4), these are probably younger clusters, just
emerging from their parental molecular cloud.

Extending out from the central clusters of Class II objects are more
filamentary groupings of stars. In W5 West, a chain of young stars
appears to extend from the cometary bright rimmed clouds at the
Southern rim of the bubble \citep[object 11 in the survey
of][]{sugitani91} up to the clusters around HD 17505 and HD 17520, and
through the cluster around BD +60 586 to the bright rimmed clouds on
the Eastern rim of the bubble. In W5 East, a chain of young stars
extends East-West from each of the two peaks of AFGL4029
\citep{price83} through the cluster around HD 18326 to the bright PAH
ridge that marks the Western edge of the bubble.

Numerous protostars (Class I objects) are seen in projection on the
cloud rim at the PAH emission/H\,{\sc ii} region cavity
boundaries. For example, the bright rimmed clouds that make up the
eastern border of W5 East, the bright rimmed clouds at the
North-Eastern rim of W5 West, and the cometary clouds at the southern
rim of W5 West all contain several Class I objects. Several isolated
cloud remnants and small `elephant trunk' formations within the
H\,{\sc ii} region cavity also contain Class I or Class II
objects---these are listed in Table 4. We give a `type' to each object
to distinguish the different morphologies. These objects are
interesting candidates for triggered star formation on small scales,
see the discussion in $\S$ 4.3. Figure 8 shows several examples of
these objects.

\begin{deluxetable}{lllr}
\tablenum{4}
\tablewidth{0pt} 
\tabletypesize{\scriptsize}
\tablecaption{Star forming globules in W5}
\tablehead{ \colhead{Object} & \colhead{RA (deg)} & \colhead{Dec. (deg)} & \colhead{Type}}
\startdata 
1 & 2$^h$ 49$^m$ 55$\fs$78 & 60$\degr$ 26$\arcmin$ 24$\farcs$46 & comet \\
2 & 2$^h$ 50$^m$ 28$\fs$16 & 60$\degr$ 09$\arcmin$ 30$\farcs$44 & comet \\
3 & 2$^h$ 50$^m$ 33$\fs$85 & 60$\degr$ 25$\arcmin$ 47$\farcs$12 & comet \\
4 & 2$^h$ 51$^m$ 32$\fs$90 & 60$\degr$ 03$\arcmin$ 53$\farcs$34 & trunk \\
5 & 2$^h$ 51$^m$ 53$\fs$91 & 60$\degr$ 06$\arcmin$ 56$\farcs$97 & comet \\
6 & 2$^h$ 52$^m$ 17$\fs$19 & 60$\degr$ 03$\arcmin$ 17$\farcs$22 & trunk \\
7 & 2$^h$ 52$^m$ 21$\fs$75 & 60$\degr$ 54$\arcmin$ 11$\farcs$86 & trunk \\
8 & 2$^h$ 53$^m$ 39$\fs$14 & 60$\degr$ 46$\arcmin$ 50$\farcs$51 & trunk \\
9 & 2$^h$ 53$^m$ 39$\fs$22 & 60$\degr$ 30$\arcmin$ 17$\farcs$95 & comet \\
10 & 2$^h$ 54$^m$ 52$\fs$34 & 60$\degr$ 35$\arcmin$ 43$\farcs$16 & trunk \\
11 & 2$^h$ 54$^m$ 58$\fs$35 & 60$\degr$ 41$\arcmin$ 43$\farcs$76 & trunk \\
12 & 2$^h$ 56$^m$  8$\fs$69 & 60$\degr$ 10$\arcmin$ 25$\farcs$26 & trunk \\
13 & 2$^h$ 58$^m$ 37$\fs$57 & 60$\degr$ 41$\arcmin$ 53$\farcs$69 & comet \\
14 & 2$^h$ 59$^m$ 46$\fs$22 & 60$\degr$ 21$\arcmin$ 10$\farcs$32 & comet \\
15 & 3$^h$  0$^m$ 23$\fs$53 & 60$\degr$ 17$\arcmin$ 55$\farcs$05 & trunk \\
16 & 3$^h$  1$^m$  1$\fs$98 & 60$\degr$ 21$\arcmin$ 57$\farcs$54 & trunk \\
17 & 3$^h$  1$^m$ 17$\fs$47 & 60$\degr$ 24$\arcmin$ 13$\farcs$20 & trunk \\
18 & 3$^h$  1$^m$ 47$\fs$94 & 60$\degr$ 35$\arcmin$ 24$\farcs$03 & trunk 
\enddata
\tablecomments{Coordinates are J2000.0.}
\end{deluxetable}

\begin{figure}
\begin{center}
\includegraphics*[width=5.5in]{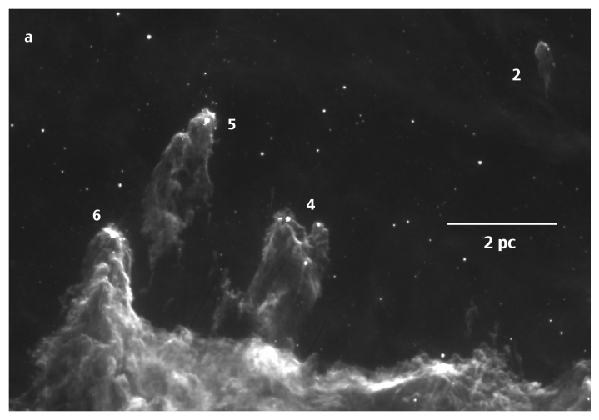}
\includegraphics*[width=5.5in]{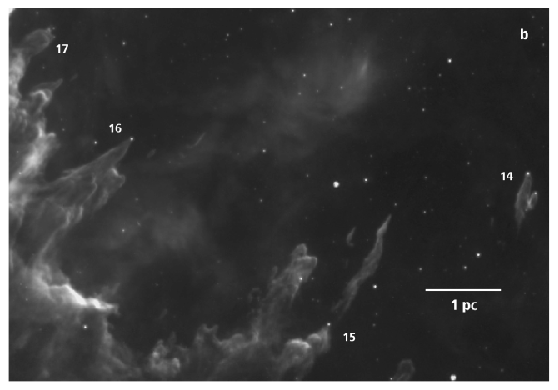}
\end{center}
\end{figure}

\begin{figure}[ht]
\begin{center}
\includegraphics*[width=3in]{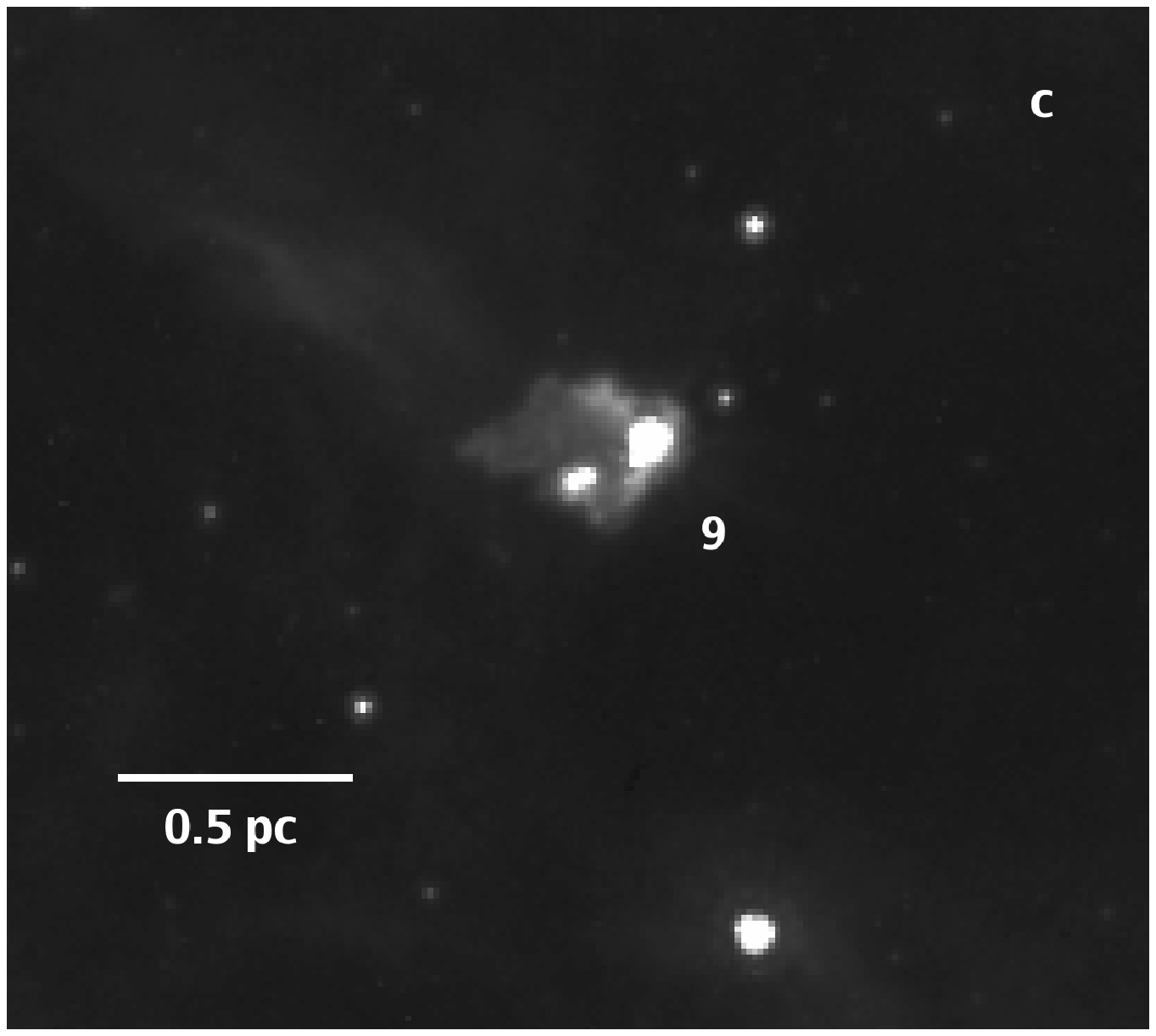}
\caption{Examples of star forming globules in W5 shown at 8
  $\micron$. PAH emission is seen at or near the globule
  heads. Objects are labelled as in Table 4.}
\end{center}
\end{figure}

The influence of star formation on its environment is clear to see in
both the IRAC (Figure 2) and MIPS images (Figures 3 and 4). Bright
diffuse emission in the four IRAC filters \citep{draine07} is only
present around the edge of the two main H\,{\sc ii} region bubbles in
roughly ring-like structures, and brightly lit-up over the molecular
clouds W5NE and W5NW. Presumably the PAHs responsible for much of the
emission in IRAC bands 1, 3 and 4 are destroyed in the harsh ionizing
environment of the H\,{\sc ii} region. At 24 $\micron$ the morphology
is similar in the bubbles' rims, but very different diffuse emission
is seen in the interior surrounding the 5 O stars (as labelled in
Figure 3). Seen in numerous massive star forming regions by
\citet{church06}, \citet{smith07} and recently \citet{harvey08}, this
additional component of diffuse emission is consistent with the
picture of \citet{crete99} whereby emission within H\,{\sc ii} regions
at mid-IR wavelengths is produced by heated larger dust grains (radius
$>$ 12 $\textrm{\AA}$) that survive being destroyed by ionizing
radiation. As \citet{draine07} have shown, emission in IRAC bands 1
and 4 is much more sensitive to the PAH mass fraction than is emission
in the MIPS 24$\micron$ band. Further, 24 $\micron$ band emission
increases rapidly when the interstellar radiation field (ISRF) rises
to 10$^4$--10$^5$ times the local value. Thus a region with low PAH
mass fraction but high ISRF can have enhanced 24 $\micron$ emission,
but will have low IRAC 3.6 and 8.0 $\micron$ PAH emission. The
morphology of the 24 $\micron$ emission suggests that these larger
grains in the {\it immediate} vicinity of the O stars are blown away
by the strong stellar winds.

\subsection{Spatial Variation in SED types}
Class I and II objects are present in varying amounts across
W5. Although with our present dataset we cannot tell if any one Class
I object is younger than a given Class II object, these classes do
represent different evolutionary stages in the formation of stars. In
W5, we find that in regions where our $^{12}$CO emission map exceeds
T$_A^*$ = 7.5 K km s$^{-1}$, there are on average 0.23 Class I objects
for every Class II. In regions with T$_A^* <$ 7.5 K km s$^{-1}$, the
ratio is 0.04. This suggests that the stellar clusters associated with
significant molecular material in W5 are systematically younger than
those in cleared out regions, for example the H\,{\sc ii} region
cavities.

The varying levels of completeness in each IRAC map as a function of
association with the bright, diffuse emission affect our result. As
described in $\S$ 2, the 90\% limiting magnitude decreases by
$\sim$1.5 mag in IRAC channels 1 and 2, by $\sim$2.5 mag in channel 3
and by $\sim$3.2 mag in channel 4 in going from the evacuated H\,{\sc
  ii} region to the brightest regions of diffuse emission. Recently,
\citet{chavarria08} studied the effects of reduced sensitivity in the
5.8 and 8 $\micron$ {\it Spitzer} bands, combined with the obscuring
effect of bright, diffuse PAH emission on the detection of Class I and
Class II sources in IRAC surveys. They found that Class II sources are
on average fainter than Class I's, and are likely preferentially
missed in bright PAH emission regions as a result.

\subsection{Clustered versus Distributed Star Formation in W5}
We want to describe in an objective way how the stars in W5 are
grouped and address the following question: is there a population of
stars that do not belong to identifiable groups that may have formed
in isolation, in other words a `distributed' population? Lacking a
dataset that describes the masses and dynamics of the stars and gas
which would allow us to assign stars to clusters based on
gravitational association, we base cluster membership {\it only} on
spatial arrangement on the sky and so we do not require that these
clusters be gravitationally bound. We restrict ourselves to infrared
excess sources only to simplify the issue of whether stars are
associated with W5. We include the deeply embedded sources and
transition disk candidates along with the Class I and Class II objects
and exclude stars exhibiting only photospheric emission (Class III
sources), AGN and PAH galaxies for a total of 2064 objects. The source
distribution is shown in Figure 7, and reproduced in Figure 9, without
the background image.

\begin{figure}
\begin{center}
\includegraphics[angle=90,width=5.4in]{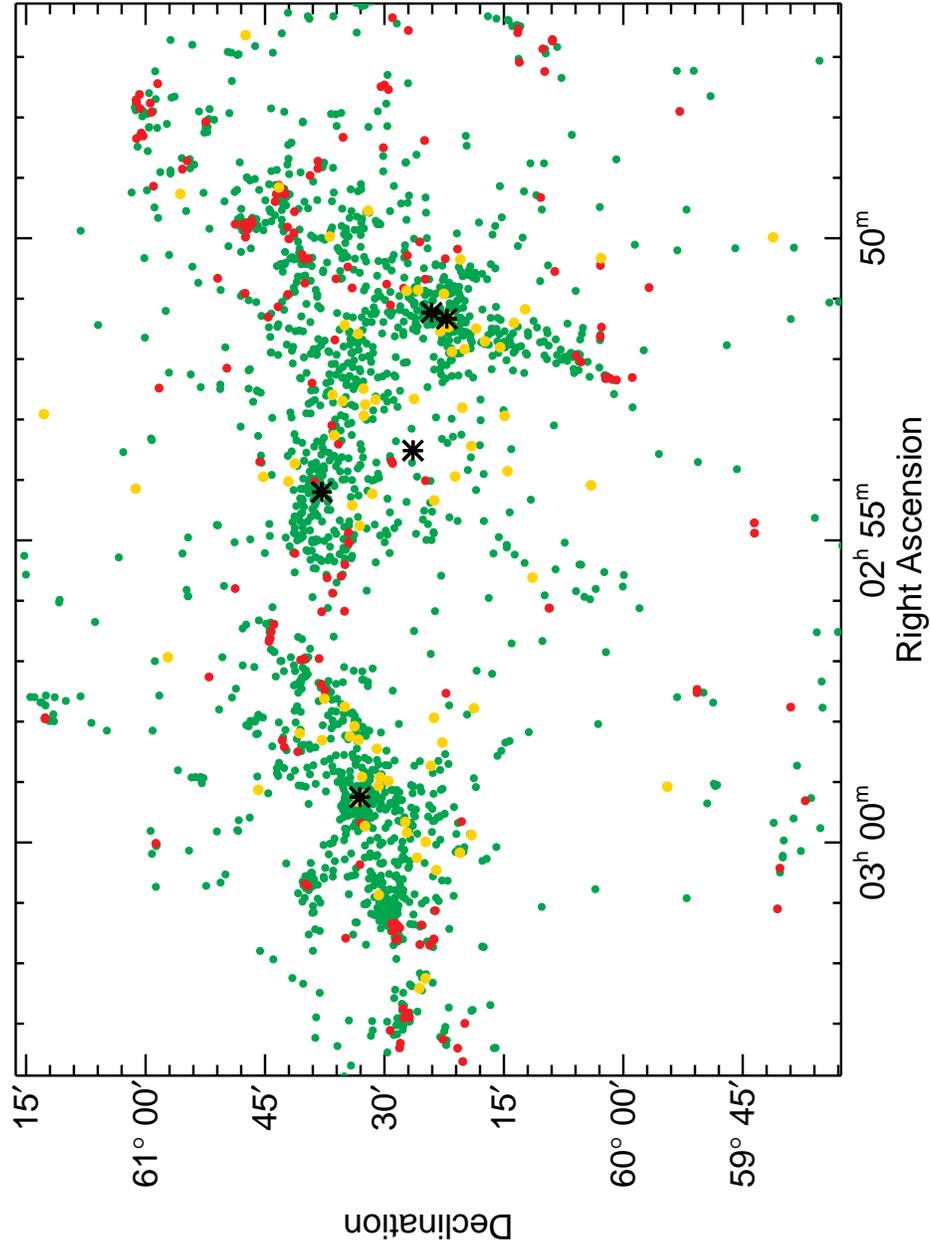}
\caption{Distribution of stellar sources as in Figure 7. Green points:
  Class II, red: Class I, yellow: transition disk candidates. Black
  asterisks mark the location of the known O stars.}
\end{center}
\end{figure}

Determining a criterion for cluster membership from spatial
distributions alone is somewhat arbitrary. In this paper we use the
so-called `minimal spanning tree' (MST) method to identify and
characterize clustering in W5 \citep[see recent work
by][]{cartwright04,bastian07}. \citet{gower69} describe how such a
tree is constructed from a list of source positions. To construct the
tree, we first generate the network of lines that joins together the
positions of all objects in our input list, such that the total length
of all lines joining points is minimized and there are no closed
loops. Each object is assigned one `branch length,' that is: the
projected distance to its nearest neighbor. We plot the distribution
of branch lengths in Figure 10.

\begin{figure}[ht]
\begin{center}
\includegraphics[width=4.0in]{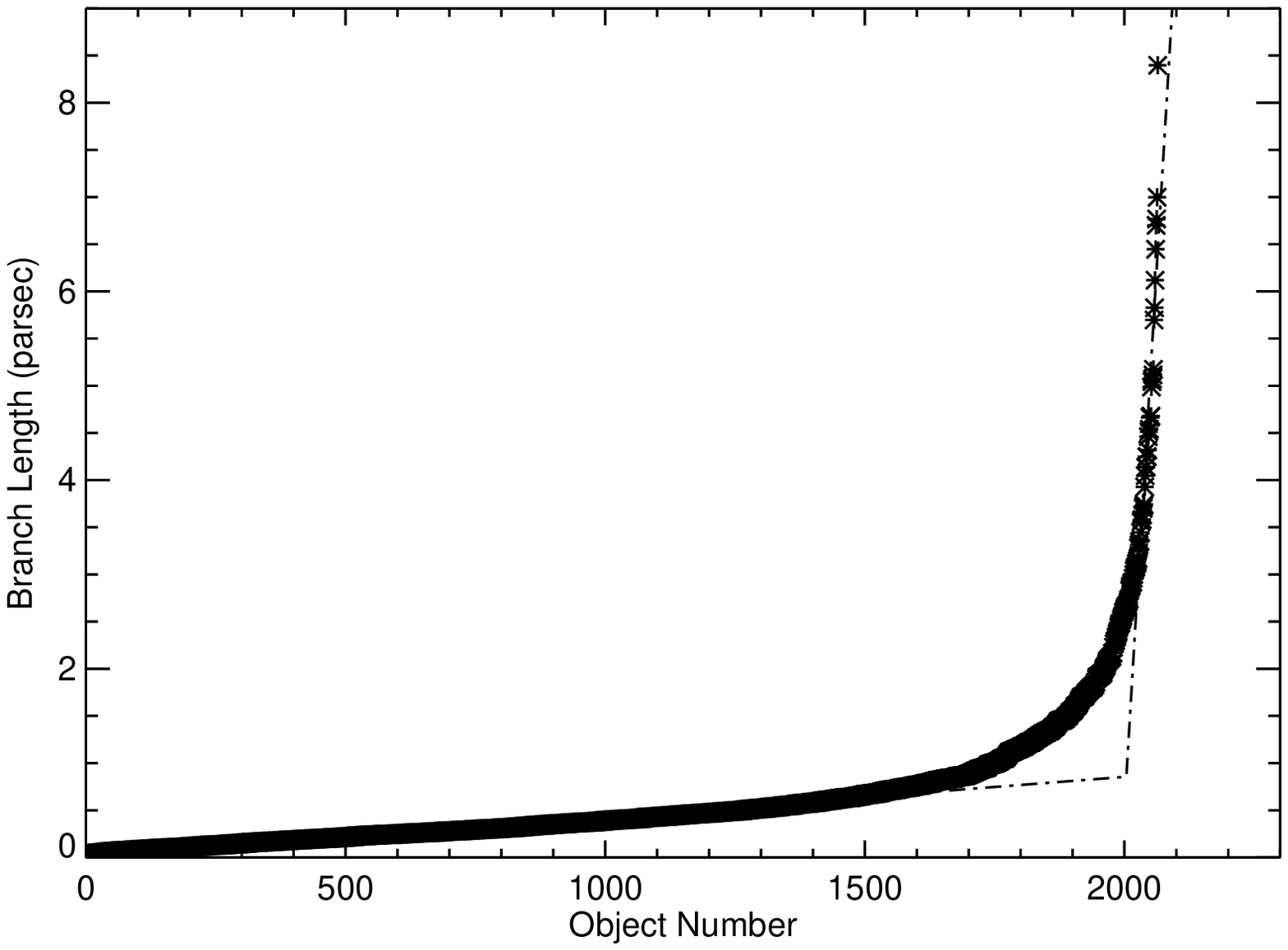}
\caption{Minimal spanning tree length distribution, showing each
  object's branch length (y-axis) versus its number in the sorted
  list. Straight line fits (dashed lines) to the short and long length
  parts of the distribution intersect at 0.86 pc.}
\end{center}
\end{figure}

To distinguish `clustered' objects from `distributed' using the MST
requires 2 things: a break/cutoff length, $d_s$ and a minimum group
size $N$. A group is then a collection of objects linked by branches
shorter than $d_s$ with at least $N$ stars (we choose $N$ = 10). We
can estimate a value for $d_s$ from the distribution of branch lengths
in Figure 10. Following Gutermuth et al. (2008, in prep) we fit
straight lines through the long and short-length portions of the
distribution. Where these lines cross defines the break length for
clustering. For the IR excess sources in W5 this method yields $d_s$ =
0.86 pc. Using this break length we find 16 groups containing at least
10 stars and a clustered fraction of 70\%.

A single value of $d_s$ chosen in this way may not be representative
of the varying spatial density of stars across the region, and may
underestimate the amount of substructure within groups. Altering $d_s$
changes the number of groups and the relative fraction of clustered
and distributed sources. As the cutoff length is increased, an
increasing fraction of objects qualify as `clustered.' The number of
groups ($N_{grp}$) we find also increases up to a maximum before
falling off as groups start to merge until all groups belong to a
single `cluster.' These trends are shown in Figure 11. An alternate
value for the break length $d_s$ can be derived from the peak in the
plot of $N_{grp}$ versus increasing $d_s$. As argued by
\citet{batt91}, this method can be thought of as returning a maximum
of information from our source distribution. In this case we find a
cutoff of $d_s$ = 0.54 pc, $N_{grp}$ = 27, and a clustered fraction of
44.2\%. In Figure 12 we compare the groupings produced by the two
methods side by side to demonstrate this.

Both values of $d_s$ used here are large when compared to the typical
separations of stars in nearby regions such as Taurus
\citep{hartmann05}, where a cutoff length of 0.54 pc would incorporate
many objects not associated with any cluster. However, Taurus is a
low-mass star forming region and so the characteristic scales of
clustering are likely to be smaller there. In recent surveys of low
mass star forming regions \citep{enoch07} and the high mass star
forming regions in Cygnus \citep{motte07}, typical star-forming dense
cores are $\sim$0.1 pc in size. However, \citep{motte07} also found
that massive dense {\it clumps} can be much larger: 0.5--0.8 pc, which
may be a more relevant scale for cluster and massive star formation
than for low density, more isolated star formation seen in Taurus.

\begin{figure}
\begin{center}
\includegraphics*[width=4.5in]{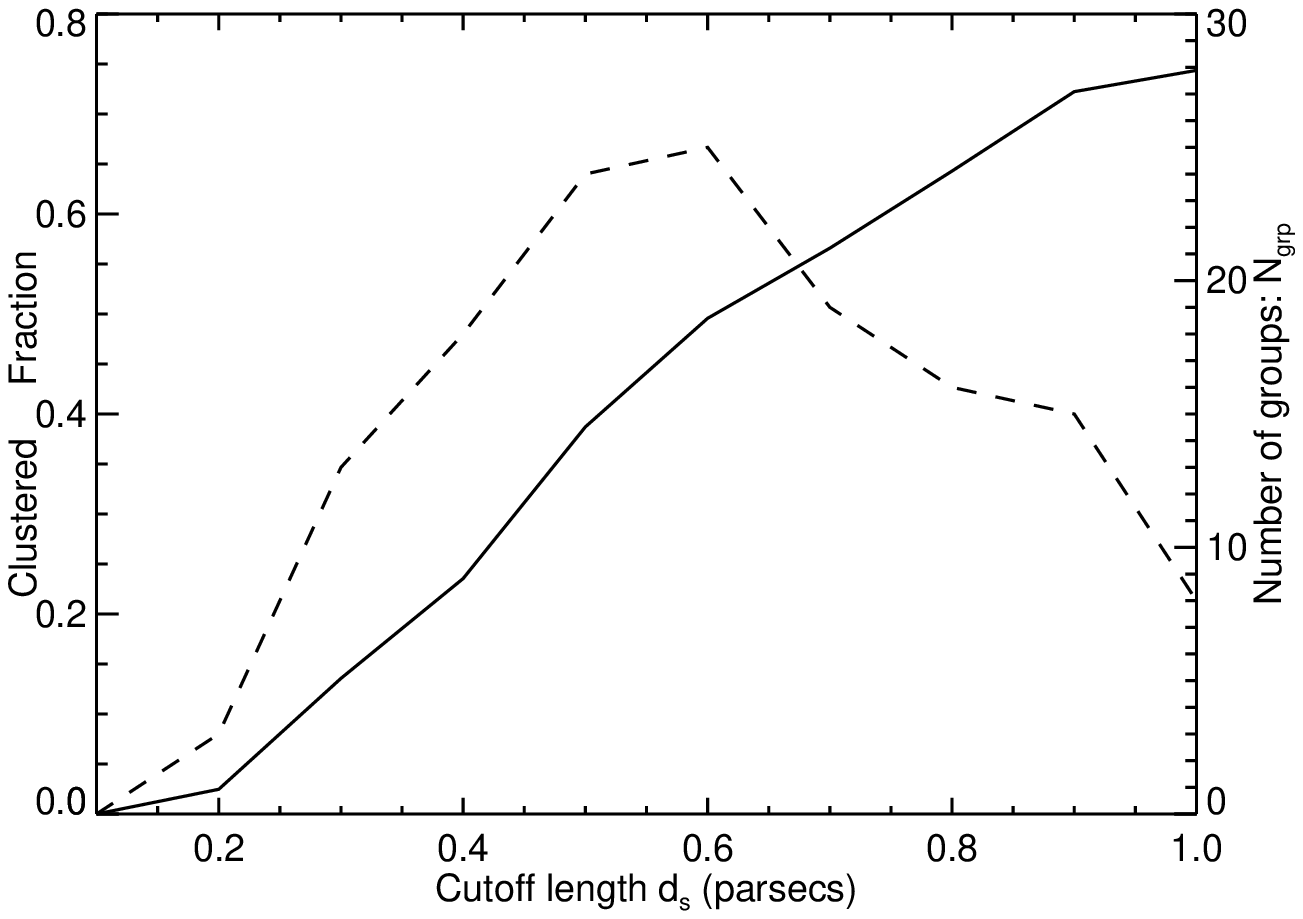}
\caption{Minimal spanning tree results. Solid line and left y-scale:
  clustered fraction of stars (belonging to all groups $\geq$10 stars)
  as a function of branch length cutoff. Dashed line and right
  y-scale: number of groups ($N_{grp}$) containing 10 or more stars
  identified by the MST algorithm. NOTE: coarse steps in x-axis miss
  finer detail in $N_{grp}$; true peak is 27 groups at $d_s$ = 0.54
  pc.}
\end{center}
\end{figure}

\begin{figure}
\begin{center}
\includegraphics*[width=5.0in]{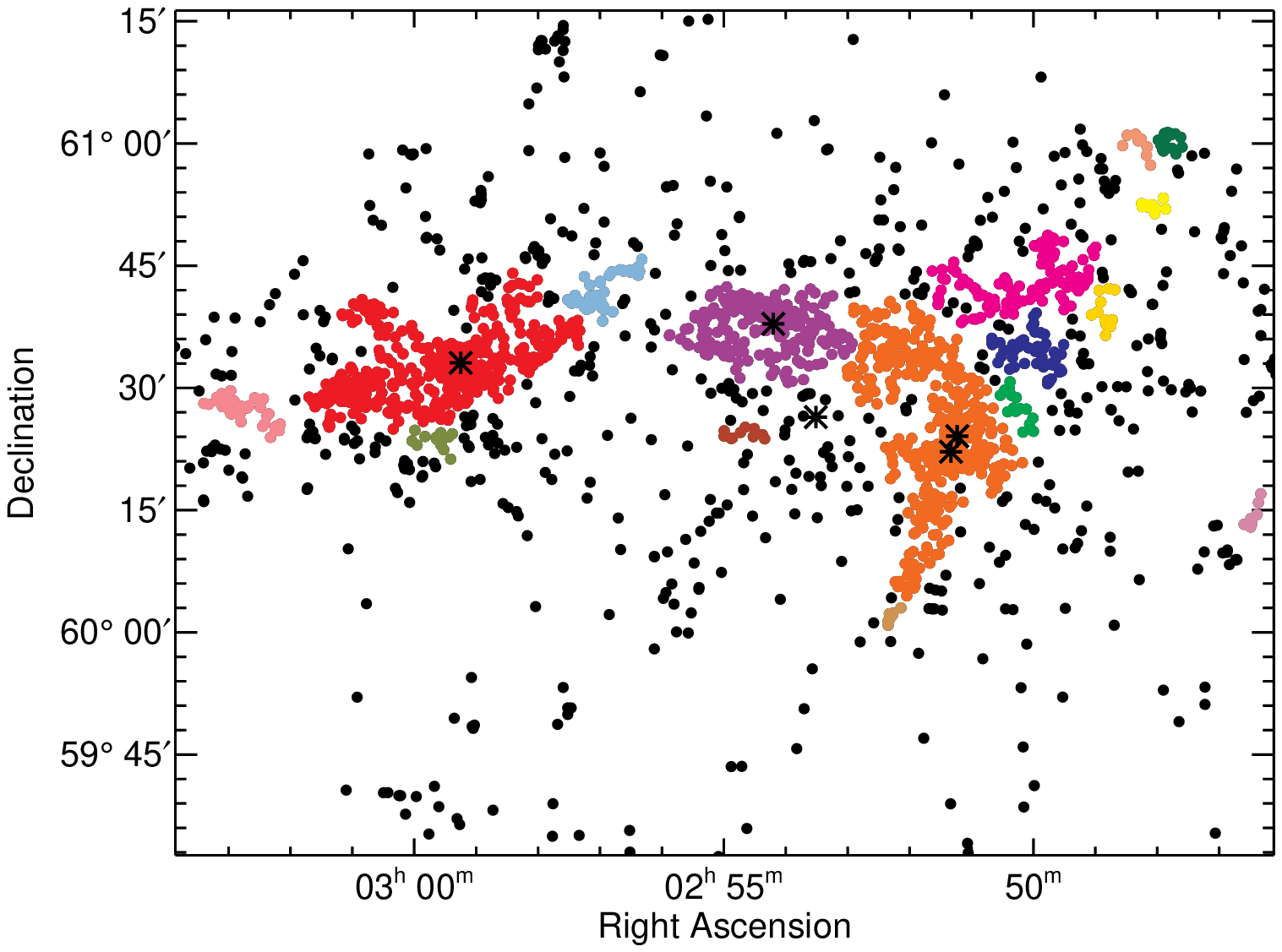}
\includegraphics*[width=5.0in]{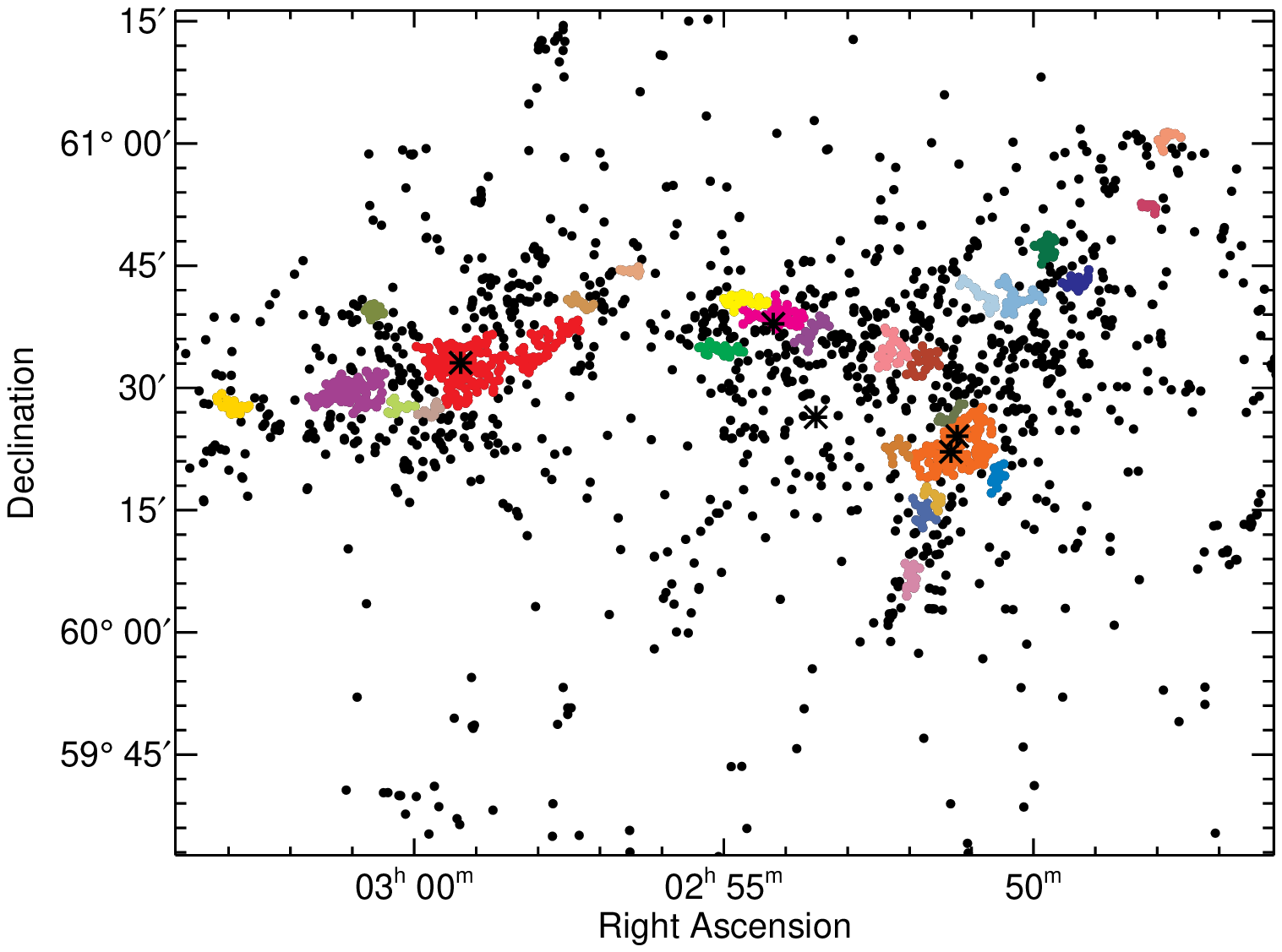}
\caption{Upper panel: the MST groups as found using the straight-line
  fit technique: $d_s$ = 0.86 pc, $N_{grp}$ = 16. Stars are colored
  according to their group for visual identification only. Lower
  panel: MST groups picked out at the maximum in $N_{grp}$ vs. $d_s$
  plot, using $d_s$ = 0.54 pc, $N_{grp}$ = 27. In both plots, black
  dots represent stars not associated with any cluster. Black
  asterisks mark the location of the known O stars.}
\end{center}
\end{figure}

Table 5 summarizes the clustering results from the two methods. The
value for the fraction of stars in clusters from both these methods is
lower than that reported by \citet{allen07} in {\it Spitzer} surveys
of Orion and Ophiuchus (74--78\%). \citet{carpenter00}, in his
analysis of the Orion A and B, Mon R2 and Perseus molecular clouds
with 2MASS found a range of clustering fractions, different for every
cloud, between 56--100\%. This fraction could be as much as a factor
two lower depending on how the age of the distributed population is
modeled. In W5---at a distance of 2 kpc---we are sampling much less of
the IMF than in these nearby regions, which lie within 1 kpc of the
Sun. We also do not know the disk-fraction of YSOs in the W5
clusters. \citet{cieza07} have shown that a significant fraction of
stars lose detectable circumstellar disks (in the mid-IR) on a
timescale $\sim$1 Myr. These objects would not be included in our
sample of IR-excess sources.

The mass completeness limit for our survey is dictated by our
photometric completeness and the wide span of YSO mid-IR colors. The
primary constraint in completeness for the IRAC 4-band sample is the 8
$\micron$ detection limit, since this has low sensitivity and very
bright background contamination from PAH feature emission. A typical
photosphere with $[3.6]-[8.0]\sim$0.1 at age 2 Myr, and $[8.0]$ =
12.7, equates to a limiting mass $\approx$2 M$_{\sun}$ (using the
evolutionary models of D'Antona \& Mazzitelli 1994, and adopting $A_V$
= 2, a typical value for the H\,{\sc ii} region cavity: Hillwig
et~al. 2006). On the bright diffuse emission the 8 $\micron$ detection
limit for a photosphere equates to $\sim$8 M$_{\sun}$. The additional
2MASS+IRAC bands 1 \& 2 sample is limited by the $K_S$ band
completeness. We estimate that the 90\% completeness in this band is
14.2. Using the models of \citet{baraffe98}, this equates to a star of
1 M$_{\sun}$ at 2--2.5 Myr, at an $A_V$ of 2 and a distance of 2
kpc. Both the 8 $\micron$ and $K_S$ band mass completeness limits are
conservative estimates, since YSOs with infrared excess emission will
be detected more readily. For example, the $[3.6]-[8.0]$ color peaks
at $\sim$1.4 for IR excess YSOs in W5. In this case, $[8.0]$ = 12.7
converts to a mass of $\approx$0.8 M$_{\sun}$ at 2 Myr in the
cavity. In summary, we estimate that for a young population of 2 Myr
age, we are complete to $\sim$8 M$_{\sun}$ for photospheres and
$\sim$4 M$_{\sun}$ for typical disk excess objects in the IRAC 4-band
sample seen against the bright background. In the cavity we are
complete to $\sim$2 M$_{\sun}$ for photospheres and $\sim$0.8
M$_{\sun}$ for disk excess objects. In the additional 2MASS sample we
are complete to a mass $\sim$1 M$_{\sun}$.

\begin{deluxetable}{lcc}
\tablecolumns{3}
\tablenum{5}
\tablewidth{0pt} 
\tabletypesize{\scriptsize}
\tablecaption{Clusters/Groups in W5}
\tablehead{ \colhead{} & \multicolumn{2}{c}{Method}  \\
\colhead{} & \colhead{Straight-line fit} & \colhead{$N_{grp}$ maximum} }
\startdata
Total $N_{\star}$ in clusters & 1444 & 912 \\
Group size ($N_{\star}$) & 10--459 & 10--201 \\
\% in clusters & 70.0 & 44.2 \\
$N_I/N_{II}$ & 0--50\% & 0--50\%
\enddata
\end{deluxetable}

%\begin{figure}
%\begin{center}
%\includegraphics[width=3.5in]{clustering_result.eps}
%\caption{Breakdown of clustering results by group size. The bins have height 0.33, 0.09, 0.04, 0.55.}
%\end{center}
%\end{figure}

\section{Discussion}
\subsection{Origin of the Distributed Population}
Our clustering analysis shows that there is a significant distributed
population of young stars (30--57\% of IR excess objects).  These may
have formed in groups or clusters and dispersed through random motions
or been ejected at high velocity.  Conversely, they may have formed in
their current locations in largely unrelated, isolated events.

On a $\sim$10 Myr timescale, the distributed population may be a
natural consequence of dynamical interactions between young stars in
groups throughout the original giant molecular cloud.  Small groups
({\it N} $<$ 36) have short relaxation times relative to their
crossing times and, once the gas is removed, will not remain visible
as clusters for much longer than this \citep{adams01}.  However, if we
assume that the distributed stars are only as old as the oldest O star
in W5 (a reasonable assumption, given that we have identified them
through their IR excess emission), they will have had $<$ 5 Myr to
travel.

Extreme dynamical interactions, namely close encounters between stars
in multiple systems, can produce high velocity runaway stars.
Numerical studies show that a given triple system will eject a member
within about 100 crossing times \citep[$\sim$30000
years:][]{reipurth00}. \citet{sterzik95} calculate similar timescales
for close systems consisting of 5 stars, and predict velocities of
3--4 km s$^{-1}$ for the ejected stars.  Few have been found;
\citet{goodman04} find 7 high velocity ($>$10 km s$^{-1}$) candidates
in the literature.

F{\H u}r{\'e}sz et al. (2006, 2008) measured the radial velocities of
young stars in NGC~2264 and the Orion Nebula Cluster, and found
dispersions of 3.5 and 3.1 km s$^{-1}$, respectively.  In W5, the
average projected distance of distributed stars from the nearest
cluster is 8.5 pc.  Assuming a velocity of 3 km s$^{-1}$ perpendicular
to the line of sight, this distance could be traversed in $\sim$3
Myr---well within the current upper age upper limit for W5 \citep[5
Myr, see:][]{karr03}. Thus it is plausible that at least some of the
distributed population originated in nearby clusters.

\subsection{Star Formation Efficiency}
We calculate the star formation efficiency in W5 using the equation:

\begin{equation}
  \epsilon = \frac{M_{stars}}{M_{stars}+M_{gas}}
\end{equation}

\noindent Here $M_{stars}$ represents the mass in {\it young} stars
only, as identified by their infrared excess, so $\epsilon$ is a
current efficiency, averaged over the last few million years. For the
entire W5 region we count 2064 young stars and assume an average mass
of 1 M$_{\sun}$. To find the gas mass we use the $A_V$ map of W5
described in $\S$ 3.1 and estimate the molecular gas column density
from the standard relation: $N(H_2)$ = 1.9 $\times$10$^{21} A_V$
protons cm$^{-2}$ \citep[see, for example:][]{bohlin78}. Assuming all
gas is associated with W5, we convert $N(H_2)$ to a projected
two-dimensional mass density at each pixel in the $A_V$ map: $\kappa$
= 15 $\times A_V$ M$_{\sun}$ pc$^{-2}$ \citep[as
in][]{lombardi01}. Each $A_V$ map pixel has a size 0.34$\times$0.34
pc. Over our whole survey field, we find a total gas mass of
6.5$\times$10$^4$ M$_{\sun}$ and derive an efficiency of 3\%.

We calculate $\epsilon$ for the clusters identified by our MST
treatment in $\S$ 3.4. For each of the 16 clusters, we count the
number of stars contained, assume a stellar mass of 1 M$_{\sun}$ and
add up the gas in the corresponding pixels. Inside the H\,{\sc ii}
region cavity, where the molecular gas has been destroyed, the
efficiencies are high (26--39\%). In the rim of W5, where molecular
gas remains and embedded clusters are found, $\epsilon$ is
10--17\%. These latter values are likely a lower limit, as our census
of young stars is incomplete in the bright extended emission
coincident with the embedded clusters.

Recent results from the c2d survey (Evans et~al. 2008 in prep.) show
that nearby dark clouds where low mass star formation is occurring
have star formation efficiencies of 2--4\% cloud-wide and 15--20\% in
clusters, similar to the values we have obtained for the high-mass
star forming region W5. Our results for $\epsilon$ in the clusters are
also comparable to those of \citet{chavarria08} for the S255 region.

%we make a 744$\times$700 grid of points
%across the W5 surveyed region. At each point in the grid we find the
%neharest 20 Class I or Class II objects and calculate their ratio. We
%convert this ratio map to a greyscale image in Figure 13 below. The
%greyscale ranges from a value of 0 (i.e. none in 20) up to 1.0 at its
%maximum near AFGL 4029. Overlaid on this image are contours of
%integrated $^{12}$CO intensity (W$_{CO}$ ranging from a minimum of 7.5
%K km s$^{-1}$ up to 157.1 K km s$^{-1}$)---a clear correlation between
%enhanced Class I numbers relative to Class II can be seen associated
%with higher CO emission. We believe this to be indicative of an age
%difference between the stellar population in the H\,{\sc ii} region
%cavity, and that associated with the molecular cloud. Note, however
%that edge effects due to low surface density of IR excess sources
%somewhat exaggerates the extent of regions of high Class I number at
%the far south-western part of W5, and in the dark region in the ??????

%\begin{figure}
%\begin{center}
%\includegraphics*[width=5.0in]{cl1cl2ratio.eps}
%\caption{Greyscale image demonstrating the variation in Class I to Class II object ratio in W5. The color scale runs from 0.0 (lightest) to 1.0 (darkest). Overlaid in black are $^{12}CO$ integrated intensity contours. Contours range from 7.5 to 157.1 K km s$^{-1}$ in uniform intervals.}
%\end{center}
%\end{figure}

\subsection{Triggered Star Formation in W5}
We consider two mechanisms of triggered star formation relevant to
W5. The first mechanism, called radiatively driven implosion (RDI), is
the compression of pre-existing density enhancements (small cloud
clumps or globules) inside and on the boundary of the ionized bubble
by the high pressure of the H\,{\sc ii} region \citep[for a review,
see][]{klein85}. Clumps of material visible as bright PAH emission in
IRAC bands 1, 3 and 4 with spatially coincident Class I or Class II
objects are summarized in Table 4. These include isolated clumps,
`elephant trunk' formations and bright rims (as shown in Figure 8). As
discussed in \citet{elmegreen98}, if globule squeezing does occur, its
timescale is expected to be short, since an isolated overdensity can
be compressed immediately on being engulfed by the H\,{\sc ii} region
or stellar wind. \citet{thompson04} investigated the cometary features
on the southern rim of W5 West for signs of star formation triggered
via the RDI mechanism (globules 4--6 in our Table 4). They found that
the H$\alpha$, CO-molecular and dust morphologies of the three pillars
are `reasonably consistent' with the model of \citet{lefloch94} for
radiatively driven implosion, at the early collapse phase. They used
near-IR photometry to detect several candidate young stellar
objects---these we confirm through our {\it Spitzer} photometry as
Class I and Class II objects. They estimate that the timescales for
duration of UV illumination, shock crossing times across the features,
and protostar/YSO ages are all $\sim$10$^5$ yrs. Since
\citet{thompson04} were only able to establish that approximate
pressure equilibrium holds in the pillar heads from their data, we
need a more detailed study of the gas dynamics in these pillars to
determine if they are currently collapsing due to external pressure
from the H\,{\sc ii} region.

The second mechanism, `collect and collapse,' is the large scale
expansion of an ionized bubble, powered by an over-pressurized H\,{\sc
  ii} region and stellar winds. This expansion can drive shock fronts
into the surrounding medium, sweeping up a dense ridge or shell which
collapses into stars when a critical density of material
accumulates. \citet{whitworth94} presented an analytical treatment of
this process. A shock front forms and gathers material until it is
able to fragment and collapse to form stars, at a time $t_{frag}$ and
radius $R_{frag}$ given by:

\begin{equation}
t_{frag} = 1.56 \textrm{ Myr }a_{0.2}^{7/11}\textrm{ }L_{49}^{-1/11}\textrm{ }n_3^{-5/11}
\end{equation}
\begin{equation}
R_{frag} = 5.8 \textrm{ pc }a_{0.2}^{4/11}\textrm{ }L_{49}^{1/11}\textrm{ }n_3^{-6/11}
\end{equation}

where $a_{0.2}$ is the sound speed inside the shocked layer in units
of 0.2 km s$^{-1}$, $L_{49}$ is the central source ionizing flux in
units of 10$^{49}$ photons s$^{-1}$ and $n_3$ is the initial gas
atomic number density in units of 10$^3$ cm$^{-3}$. 

In any massive star forming region, presumably some combination of
H\,{\sc ii} region expansion and stellar wind forces operate. Stellar
winds certainly play a significant role in shaping the morphology of
an H\,{\sc ii} region only until an age of $\sim 10^5$years. As shown
by \citet{mckee84} and \citet{weaver77}, weak stellar winds (wind
luminosity $\ll$ 1.26$\times$10$^{36}\times L_{49}\times n$, where $n$
is the ambient density) are likely to be confined by the H\,{\sc ii}
region pressure to a smaller bubble around the ionizing source, since
gas that is evaporated from clumps of material in the H\,{\sc ii}
region mixes with the hot stellar wind bubble and can radiate energy
away efficiently. Strong winds can overcome this confinement---they
create a bubble with size determined by the photoevaporation and
displacement of surrounding inhomogeneities in the ambient gas. This
then expands along with the H\,{\sc ii} region.  In estimating the
lifetimes of the W5 bubbles, we thus consider only the effects of the
H\,{\sc ii} region expansion.

As described in $\S$1, the gross morphology of W5 in the mid-IR is
defined by two large, roughly circular rings of bright PAH emission
that mark the smaller W5 East (W5E, radius $\sim$10 pc), and larger,
more irregular W5 West (W5W, overall radius $\sim$22 pc). At the
approximate center of W5E lies HD 18326, an O7V star surrounded by a
dense cluster of young stars. It seems reasonable to assume that
whether this is a sphere or a ring of gas, that the cluster lies near
the true center, and is not significantly offset either to the
foreground or background. Adopting stellar parameters from
\citet{martins05}, an O7V star will have ionizing flux $L_{49}$ =
0.43. Since we cannot calculate appropriate values for $a_{0.2}$ and
$n_3$ from our current datasets, we adopt 1.0 for both. The H\,{\sc
  ii} region fragmentation time we calculate is 1.69 Myr at a radius
of 5.37 pc. Both \citet{whitworth94} and \citet{dale07} note that the
density in a real region is likely to be lower---this would make the
fragmentation time longer and the radius larger by a factor
$\sim$2. 

W5W presents a less clear-cut picture. It contains an isolated O8V
star: HD 237019, as well as three dense clusters centered on BD +60
586 (O7.5V), and the multiple systems HD 17505 (two O7.5V((f)) stars,
an O6.5III((f)) and an O8.5V) and HD 17520 (O9V + Be). See
\citet{hillwig06} for a more detailed study of the O stars in W5. The
relative configuration of these four systems, and their relationship
to the surrounding diffuse gas (whether in front or behind) is not
known. We consider here two simple scenarios and assume that all four
objects and the ring of PAH emission are in the same plane. Scenario
1: HD 237019 represents an initial episode of star formation, and its
isolation is due to its cluster having had time to disperse. In the
Whitworth model, an O8V star has a fragmentation time of 1.8 Myr and
collapse of swept up gas occurs at a radius of 5.0 pc. In this
picture, this event triggered the formation of the three dense
clusters in W5W which all lie at roughly 6.8 pc from HD 237019. These
clusters, and the continuing expansion of the H\,{\sc ii} region then
triggered the current, ongoing star formation seen in association with
molecular clouds W5NW and W5NE. Scenario 2: the three dense clusters
form together---with their present arc-like arrangement due to some
initial filamentary distribution of molecular material. They then
trigger star formation in the remaining molecular material to the
north, north-east and south. The O7.5V star BD +60 586 by itself has a
fragmentation time of 1.75 Myr and collapse occurs at a radius of 5.16
pc. The combined systems HD 17505 and HD 17520 \citep[with at least 5
O stars, see:][]{hillwig06} have a fragmentation time of 1.46 Myr and
collapse occurs at a radius of 6.19 pc. In this picture, HD 237019
would have been ejected from one of the multiple O star
systems. Located at $\approx$10 pc from the nearest cluster, assuming
a velocity entirely transverse to our line of sight of 10 km s$^{-1}$,
it would take $\sim$1 Myr to arrive at its current location,
consistent with the likely young age of the cluster as a
whole. \citet{moffat98} found a runaway frequency percentage of 14\%
among Galactic O stars and included this star as a marginal runaway
candidate. Although the most favored creation mechanism for runaways
is supernova ejection \citep{blaauw61}, ejection from compact young
stellar clusters or through binary interaction is also thought to be
responsible for some O star runaways
\citep{poveda67,clarke92}. \citet{moffat98} found HD 237019 to be only
a marginal runaway candidate, given the large uncertainty in its
tangential velocity: 15$\pm$17 pc Myr$^{-1}$.

There are several observations consistent with the collect and
collapse mechanism in W5. The low level of non-thermal radio emission
suggests there have been no supernovae during its lifetime
\citep{vallee79}. The main sequence lifetimes of the O stars are
estimated to be $<$ 5 Myr \citep{karr03}. The timescales for
triggering presented above are shorter than this. \citet{wilking84}
found several young embedded OB stars around W5 in the molecular
clouds, and argued that their locations near to the cloud edge,
together with some evidence of magnetic field alignment parallel to
the ionization front are suggestive of triggering via external
compression. Finally, \citet{nakano08} found evidence of an age
difference of $\sim$3 Myr between the cluster around HD 18326 in the
W5 East bubble and the young stars in the rim. This is longer than our
estimate above of $\sim$1 Myr, but given the uncertainties in gas
sound speed $a$ and initial cloud number density $n$, is an allowable
timescale in the collect and collapse model.

\citet{karr03} argue that the distribution of young stars in W5
indicates a shorter timescale, more consistent with RDI than with
collect and collapse. They investigated the locations of IRAS point
sources with colors corresponding to YSOs. Their results did suggest a
2-generation model, with older stars at the centers of the H\,{\sc ii}
regions and ongoing star formation surrounding them, however they
found that the number density of young objects peaks $\sim$5 pc inside
the H\,{\sc ii} region (as defined by the 6.2K 1420 MHz contour),
which corresponds to a triggering timescale of 0.5--1 Myr. Of the 42
IRAS sources from their list that fall within our {\it Spitzer} image
we find only 16 in our source list at 24 $\micron$. The remainder may
be spurious (they are also not seen in our 70 $\micron$ image) and in
fact may be small knots of IR emission that masquerade as point
sources in the IRAS survey. All are associated with bright PAH
emission, including one coincident with globule 8 in Table 4. The peak
in IRAS-classified YSOs found by \citet{karr03} is also not upheld by
an analysis of our {\it Spitzer} sample which shows that the Class I
objects are found predominantly along the cloud rims or at the cloud
centers. This distribution, with the YSOs further from the O stars
argues for the longer timescale of collect and collapse.

Studies of a young embedded object G138.295+1.555 and the UCH\,{\sc
  ii} region G138.300+1.558 in AFGL4029 (W5E) with near- and mid-IR
imaging \citep{deharveng97,zavagno99} suggest that collect and
collapse is not playing a role in star formation here, since the
youngest object (the former) is apparently closer to the ionization
front than the older, latter object. However the true 3-dimensional
configuration of all the objects involved is not exactly known, this
may be due to projection effects. The clustering of Class II objects
immediately outside the cloud to the west of G138.295+1.555 and
G138.300+1.558 is certainly suggestive that the sequence of star
formation here is west to east, i.e. outside-in, away from the
ionizing star HD 18326.

In consideration of our simple model and the results in the
literature, it seems plausible that both the RDI and collect and
collapse mechanisms are at work in W5; RDI on the smaller scale of
cometary globules, and collect and collapse on the larger scale of the
H\,{\sc ii} region. Detailed investigations (Koenig et al. 2008, in
preparation) of the spatial distributions and relative ages of the
YSOs may further constrain the scenarios presented here.

\section{Conclusions and Future Work}
We have presented initial results from our extensive {\it Spitzer}
survey of W5. Shorter wavelength data from IRAC (3--8 $\micron$) and
longer (24 $\micron$) wavelength data from the MIPS instrument were
combined to maximize spectral coverage of detected sources.

Even before photometric analysis, dense clusters of stars are clearly
visible across the region, centered on the O stars HD 18326, BD +60
586, HD 17505 and HD 17520, and also across the extensive PAH emission
that defines the outline of W5. At 24 $\micron$, substantial extended
emission is visible, presumably from heated dust that survives in the
strongly ionizing environment of the H\,{\sc ii} region.

We used photometry of more than 18000 point sources to analyze the
spatial distributions of young stars, establish their evolutionary
status via their infrared colors and magnitudes, and assess their
clustering properties across this large star forming region. The large
clusters that dominate the region, centered on the massive O stars,
contain numerous infrared excess sources. Looking at the large scale
distributions of stars at different evolutionary stages, we find that
within the evacuated cavity of the H\,{\sc ii} regions that make up
W5, the ratio of Class II (older) to Class I (younger) sources is
$\sim$7 times higher than for objects detected coincident with the
molecular clouds in the rim. We attribute this difference to an age
difference between the two locations, and consequently postulate that
at least two, distinct generations of star formation are visible in
the region. An isolated O star in W5 West, HD 237019, may represent an
initial episode of star formation in the region, preceding the
formation of the large clusters in the cavity, although we cannot rule
out its ejection from an O star multiple system, for example HD 17505.

The clustering results show that, considering infrared excess sources
alone, (2064 objects) $\sim$45--70\% are found in clusters with
$\geq$10 members. Incorporating the sources apparently misclassified
as AGN and PAH galaxies (see Appendix) extends this range to
$\sim$40--70\%. The remainder are in the distributed mode, many of
which could have formed in nearby clusters.

We looked at the role that {\it triggered} star formation may have
played in W5. We catalogued isolated globules of diffuse PAH emission,
and so-called `elephant-trunk' structures that contain young
protostars or stars with protostellar disks. These are examples of
possible RDI-triggered star formation events. On the larger scale, we
tested the analytical formulations of \citet{whitworth94} for the
collect and collapse mechanism. Our simple estimates show that
triggering remains a plausible mechanism to explain the multiple
generations of star formation in W5 and merits further investigation.

Substantial work is ongoing to refine our understanding of star
formation in W5 in the light of triggering models and the clustering
of stars. We have undertaken a near infrared survey of the whole
region in $J$, $H$ and $K_S$ bands to extend the stellar SEDs to
shorter wavelengths and detect more young stars against the bright
background emission in W5. This will permit a full analysis of the
clustering of young stars in the region.  We have obtained optical
spectra of several hundred stars across W5 with the aim of determining
their relative ages by constructing H-R diagrams. This will allow us
to better understand the history of star formation across the entire
region. A much clearer picture of the progression of star formation
across W5 and the role of feedback in this process should result.

\acknowledgements The authors would like to thank Paul Harvey and Phil
Myers for a very helpful analysis of the manuscript and useful
discussions, and Fred Adams, Tom Megeath and Neal Evans for helpful
discussions on clustering analysis. This work is based (in part) on
observations made with the {\it Spitzer} Space Telescope, which is
operated by the Jet Propulsion Laboratory, California Institute of
Technology under a contract with NASA. Support for this work was
provided by NASA. This publication makes use of data products from the
Two Micron All Sky Survey, which is a joint project of the University
of Massachusetts and the Infrared Processing and Analysis
Center/California Institute of Technology, funded by the National
Aeronautics and Space Administration and the National Science
Foundation. This research has made use of NASA's Astrophysics Data
System. This research has made use of the SIMBAD database, operated at
CDS, Strasbourg, France.

{\it Facilities:} \facility{2MASS ($JHK_S$)}, \facility{Spitzer (IRAC, MIPS)}

\section*{Appendix}
\appendix
\section{AGN/PAH Galaxy filtering}
Following \citet{gutermuth08} PAH galaxy candidates are identified on
the basis of the following criteria:

\begin{displaymath}
[4.5]-[5.8] < \frac{1.05}{1.2} ([5.8]-[8.0]-1),
\end{displaymath}
\begin{displaymath}
[4.5]-[5.8] < 1.05,
\end{displaymath}
\begin{displaymath}
[5.8]-[8.0] > 1,
\end{displaymath}
\begin{displaymath}
[3.6]-[5.8] < \frac{1.5}{2} ([4.5]-[8.0]-1),
\end{displaymath}
\begin{displaymath}
[3.6]-[5.8] < 1.5,
\end{displaymath}
\begin{displaymath}
[4.5]-[8.0] > 1.
\end{displaymath}

After removing these sources, AGN are then picked out following these
criteria:

\begin{displaymath}
[4.5]-[5.8] > 0.5,
\end{displaymath}
\begin{displaymath}
[4.5] > 13.5 + ([4.5]-[8.0]-2.3)/0.4,
\end{displaymath}
\begin{displaymath}
[4.5] > 13.5.
\end{displaymath}

We plot the distribution of 729 AGN and 198 PAH galaxy candidates in
W5 in Figure 13. It is notable that these objects show obvious
non-uniformity. The non-uniformity arises from several sources: 1)
variation in the intrinsic distribution of galaxies toward (behind)
W5, 2) variable extinction across the region due to the molecular
clouds that make up W5 and material along the line of sight through
the Galaxy in this direction, 3) variable levels of photometric
completeness across the field due to bright extended nebular emission
and 4) clustering coincident with the positions of the dense clusters
of young stars around the O stars in the W5 H\,{\sc ii} region cavity,
due to faint (low mass) stars misclassified as AGN or PAH
galaxies. This last issue affects all of our analysis, thus we need to
characterize---statistically---the properties of these misclassified
objects and how adding them back to the original sample affects our
previous results.

\begin{figure}
\begin{center}
\includegraphics[width=4in]{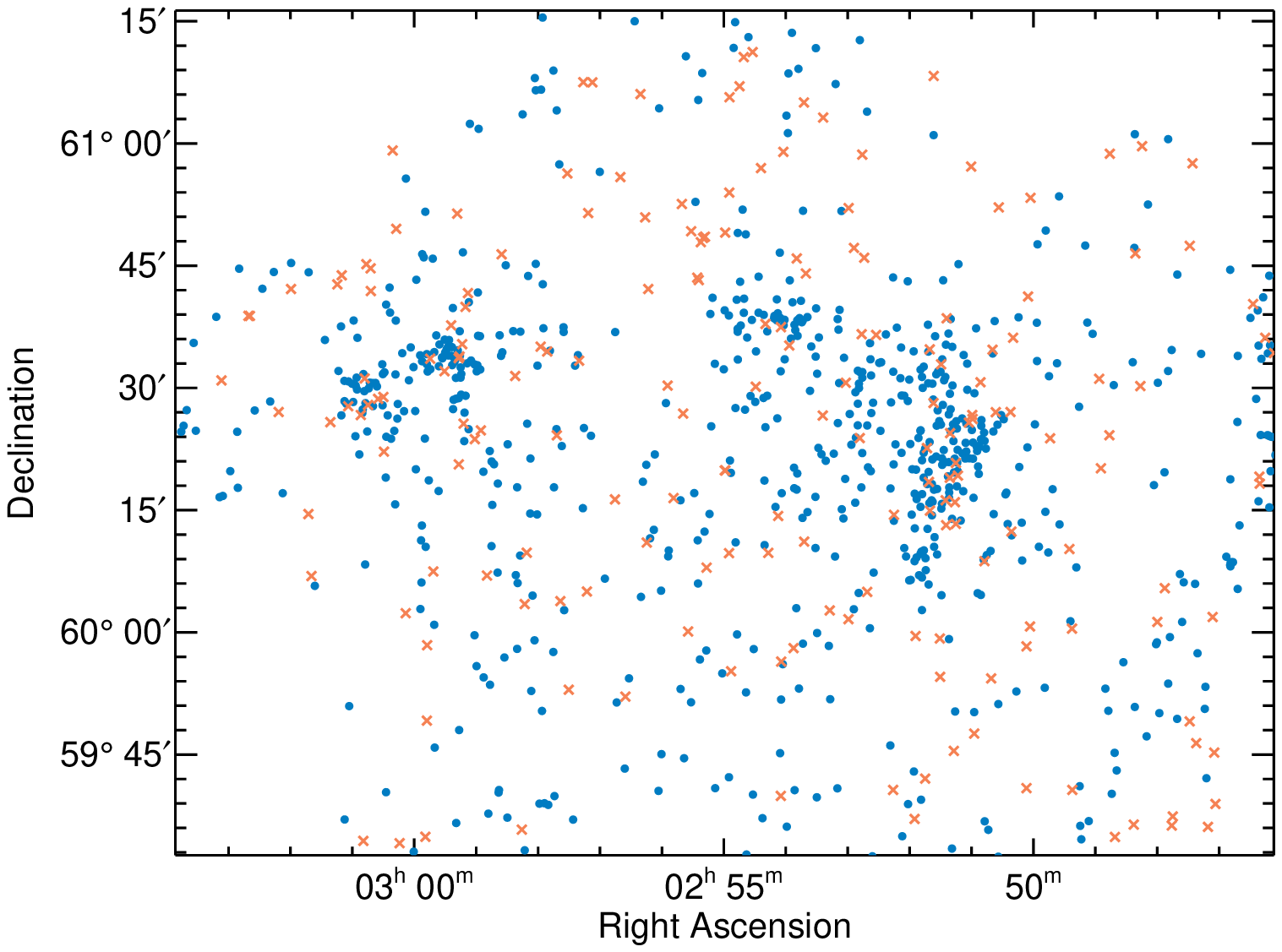}
\caption{Distribution of extragalactic contaminants as classified by {\it Spitzer} photometry: AGN (blue points) and PAH galaxies (red).}
\end{center}
\end{figure}

We consider the AGN first. In the extreme case where all objects are
misclassified, we return the full list of 729 AGN to the stellar list,
classify the returned objects into Class I, II etc., and examine the
clustering fraction. With the MST straight line fit method, we find
$d_s$ = 0.88 pc, $N_{grp}$ = 21 and a clustered fraction = 68.6\%. If
we use the maximum in the $N_{grp}$ distribution we find $d_s$ = 0.5
pc, $N_{grp}$ = 26 and a clustered fraction = 40.4\%.

Instead of this, presumably some intermediate fraction of AGN should
be returned as AGN. We must first calculate the expected number and
averaged distribution of AGN in W5, using a sample of AGN extracted by
\citet{stern05} from the {\it Spitzer} IRAC shallow survey of the
Bo\"otes field \citep{eisenhardt04}, convolved with our extinction map
(as described in $\S$ 2.1) and with completeness estimates derived for
our {\it Spitzer} survey of W5 via a simple Monte Carlo simulation
code. The model AGN sample is drawn from a 7.7 square degree section
of the \citet{stern05} survey, identifying sources as AGN based on the
same color classification scheme that we use in this paper
\citep{gutermuth08}.

We use the IRAC completeness maps as described in $\S$ 3.1. The area
imaged in our W5 survey (3.5 square degrees) is smaller than that
covered by the Bo\"otes data. In our Monte Carlo simulation, we input
a proportionately smaller, random selection of AGN from the Bo\"otes
list and distribute these randomly in the field of W5. Each object is
assigned an $A_V$ value based on its location in the extinction map.
We apply this extinction to its IRAC photometry using the infrared
extinction relation presented in \citet{flaherty07}. The corresponding
location within the completeness maps gives us the completeness value
(a number $\leq$1) at each of the four IRAC wavelengths given its
extincted magnitude. A random number generator is then used to
determine whether or not an object is detected, with the completeness
estimates as a measure of the probability of detection. For detection
we require that an object be detected in all four IRAC bands. This
simulation predicts on average 270 AGN detected in W5.

This distribution of AGN has a certain characteristic space density
distribution. At each iteration we measure the `nearest-neighbor'
distance distribution \citep{casertano85}. For each object we find the
projected distance to its sixth nearest neighbor in arcsec ($d_6$) and
construct a histogram of distances. The final output of the code is an
averaged histogram combining 1000 outcomes of the simulation. For
comparison, we generate the distribution of $d_6$ for the real AGN
candidates in W5 and compare the results in Figure 14 (leftmost
panel). In the right-hand panels we compare the output magnitude and
color distributions of the simulated AGN (averaged over 1000 outcomes)
with our W5 AGN candidates.

\begin{figure}[ht]
\begin{center}
\includegraphics*[width=2.12in]{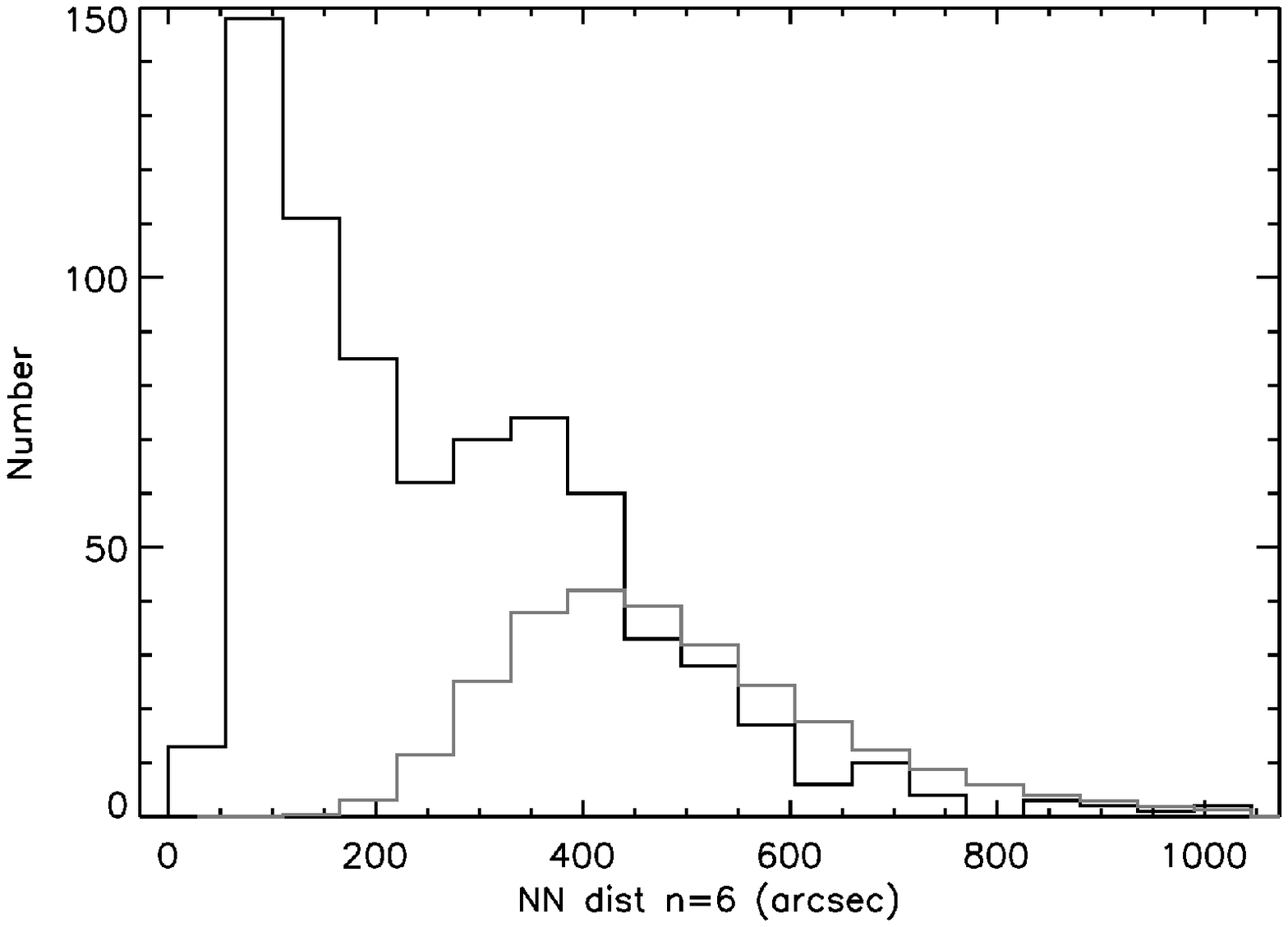}
\includegraphics*[width=2.12in]{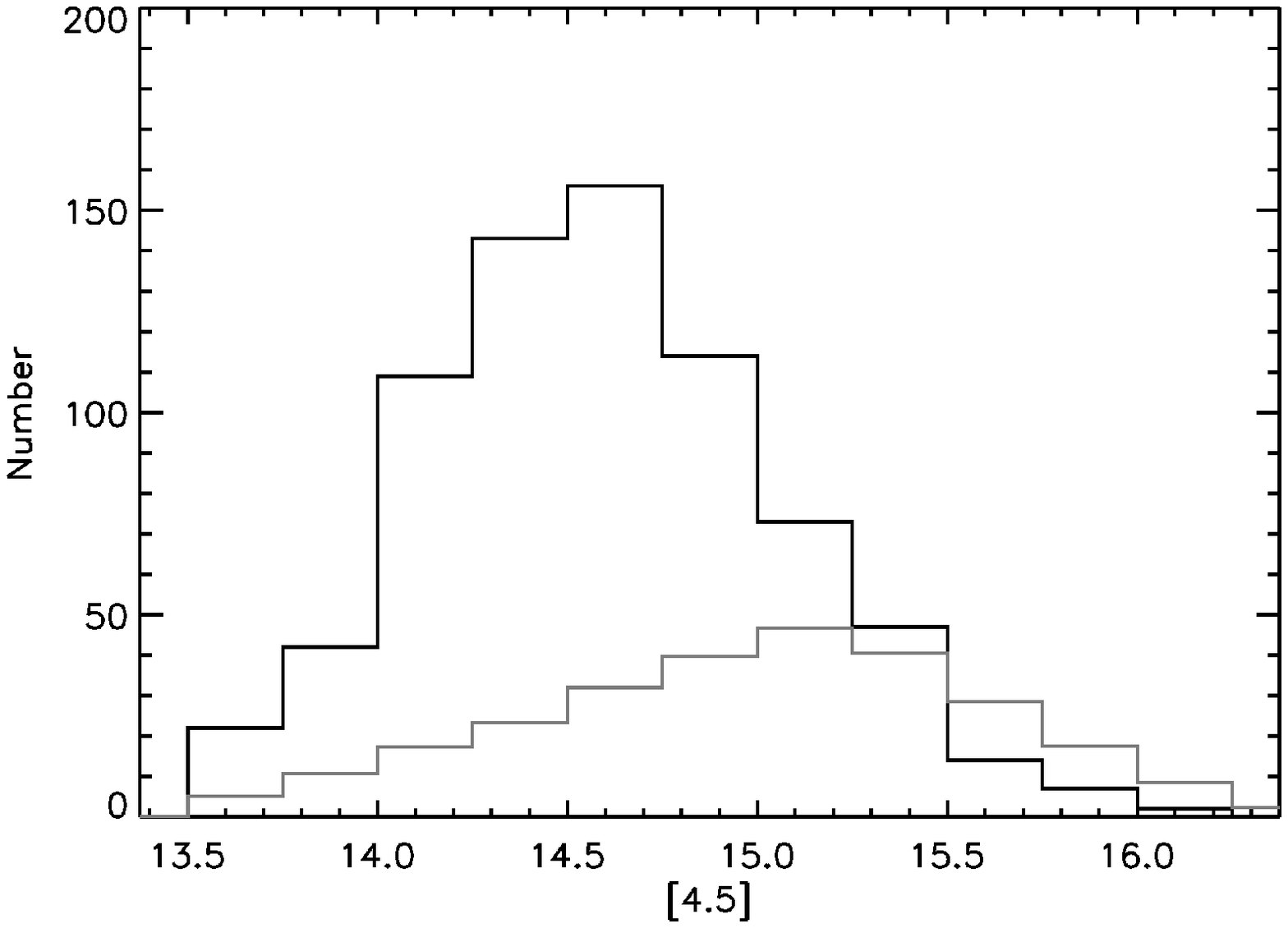}
\includegraphics*[width=2.12in]{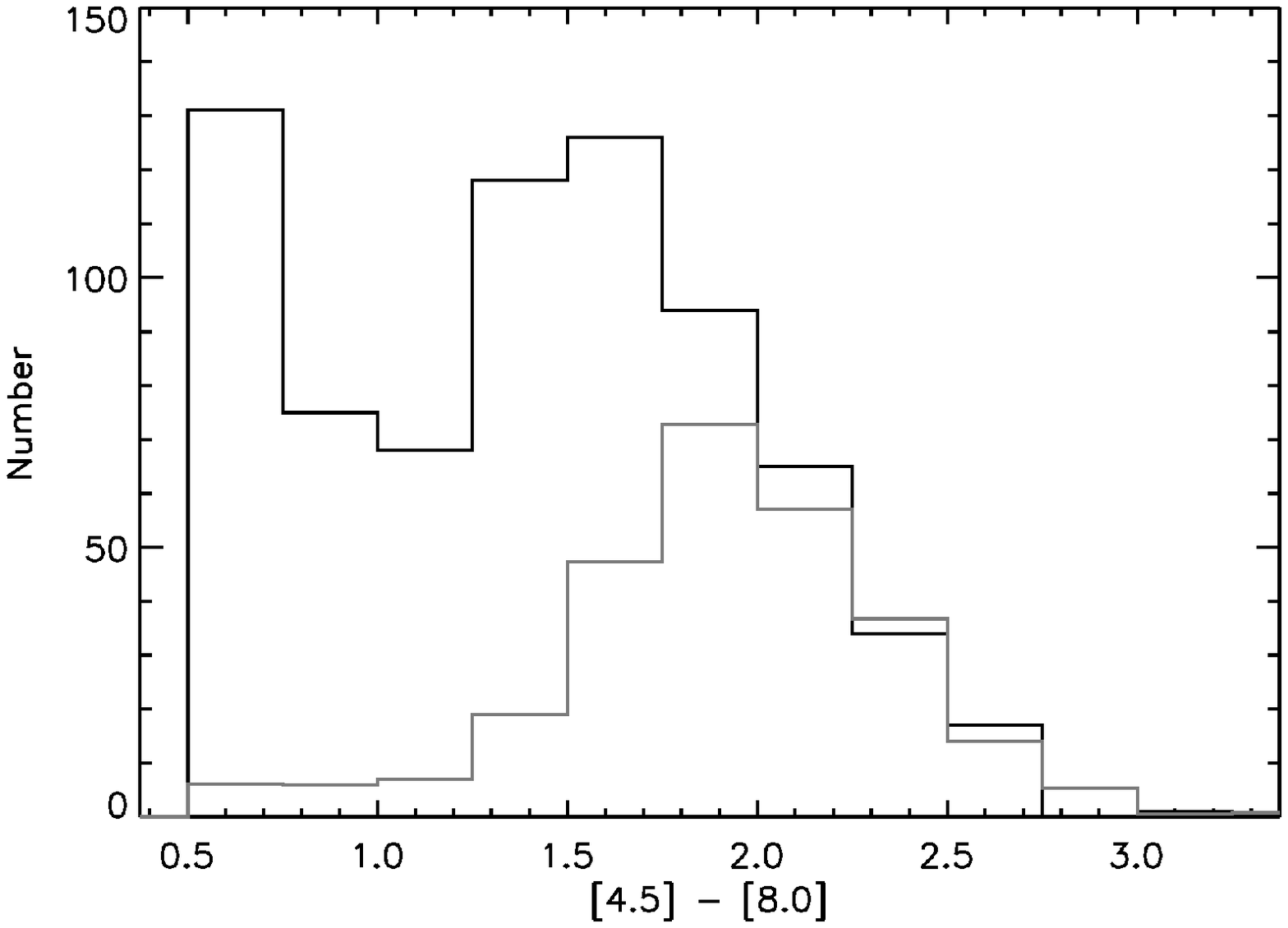}
\caption{Distributions of nearest neighbor distances, [4.5] magnitudes
  and $[4.5]-[8.0]$ colors for objects classified as AGN in W5 (black
  histograms). Plotted in gray are the simulated distributions of
  simulated AGN according to our completeness estimates. We attribute
  the excess of objects in the W5 sample to young stars misclassified
  as AGN by our classification scheme.}
\end{center}
\end{figure}

These distributions can be used to generate a probability that an
object in our AGN list, of given $d_6$, $[4.5]$ magnitude and
$[4.5]-[8.0]$ color is a YSO. In Table 6 we present the source
properties of the 729 W5 AGN candidates. The full machine-readable
table is given in the online material for this paper. Each object is
listed with its coordinates and photometry in 2MASS and {\it Spitzer}
bands, and given a combined `P(YSO)' value, that is the probability
given its local surface density, $[4.5]$ magnitude and $[4.5] - [8.0]$
color.

We statistically return a sample of AGN to the YSO list based on these
probabilities to test the effect on the clustered fraction. We return
samples based on the density probability alone, and then the combined
density, color and magnitude probability. We run our MST clustering
analysis on the resultant list of YSOs plus YSO-AGN. In Table 7 below
we show the averaged results of adding AGN back to the YSO sample
according to the different filters and compare to the original
YSO-only sample result.

\begin{deluxetable}{lcccccccccccc}
\tabletypesize{\scriptsize}
\rotate
\tablenum{6}
\tablewidth{0pt}
\tablecaption{W5 Extragalactic Sources}
\tablehead{\colhead{ } & \colhead{RA} & \colhead{Dec} & \colhead{$J$} & \colhead{$H$} & \colhead{$K_S$} & \colhead{$[3.6]$} & \colhead{$[4.5]$} & \colhead{$[5.8]$} & \colhead{$[8.0]$} & \colhead{$[24]$} & \colhead{ } & \colhead{ } \\
\colhead{ID} & \colhead{(deg)} & \colhead{(deg)} & \colhead{(mag)} & \colhead{(mag)} & \colhead{(mag)} & \colhead{(mag)} & \colhead{(mag)} & \colhead{(mag)} & \colhead{(mag)} & \colhead{(mag)} & \colhead{Type\tablenotemark{a}} & \colhead{P(YSO)\tablenotemark{b}}}
\startdata 
g1 & 41.176602 & 60.672605 & \nodata & \nodata & \nodata & 15.52(03) & 14.62(03) & 14.71(17) & 13.28(18) & \nodata & PAH & 0.00 \\
g2 & 41.189968 & 60.652874 & \nodata & \nodata & \nodata & 15.17(03) & 14.48(04) & 14.07(11) & 13.02(11) & \nodata & AGN & 0.28 \\
g3 & 41.190158 & 60.586695 & 17.00(19) & \nodata & 15.24(14) & 14.04(01) & 13.65(02) & 12.98(06) & 11.99(07) & \nodata & AGN & 0.22 \\
g4 & 41.242231 & 60.703623 & \nodata & \nodata & \nodata & 14.72(03) & 13.85(04) & 13.31(12) & 12.19(17) & \nodata & AGN & 0.05 \\
g5 & 41.245464 & 60.736165 & \nodata & \nodata & \nodata & 14.71(03) & 13.93(03) & 13.35(08) & 12.38(07) & \nodata & AGN & 0.10 \\
g6 & 41.252453 & 60.541187 & \nodata & \nodata & \nodata & 14.53(03) & 13.97(02) & 13.48(07) & 12.37(06) & \nodata & AGN & 0.37 \\
g7 & 41.256135 & 60.465426 & 14.27(04) & 13.83(04) & 13.62(04) & 13.47(01) & 13.51(02) & 13.57(08) & 11.77(03) & \nodata & PAH & 0.02 \\
g8 & 41.274375 & 60.732626 & 16.90(17) & 16.23(24) & 15.34(15) & 14.65(03) & 14.15(03) & 13.71(12) & 11.95(08) & \nodata & PAH & 0.00 \\
g9 & 41.293823 & 60.525857 & 15.26(05) & 14.61(06) & 14.55(08) & 14.36(01) & 14.32(03) & 14.25(12) & 13.69(18) & \nodata & AGN & 0.63 \\
g10 & 41.294108 & 60.520992 & \nodata & \nodata & \nodata & 15.30(03) & 14.64(03) & 13.96(09) & 12.96(09) & \nodata & AGN & 0.40 \\
\enddata
\tablecomments{Table 6 is published in its entirety in the electronic edition of the {\it Astrophysical Journal}. A portion is shown here for guidance regarding its form and content. Values in parentheses signify error in last 2 digits of magnitude value. Right ascension and Declination coordinates are J2000.0.}
\tablenotetext{a}{Galaxy type: AGN or `PAH galaxy' as defined in text.}
\tablenotetext{b}{Probability that source is a YSO based on local space density, color, magnitude criteria: see text.}
\end{deluxetable}

The space density filter typically returns $\approx$460 objects to the
list. The combined filter returns $\sim$230.  Although a significant
fraction of objects that are returned from AGN to YSOs join the
clustered population, many are added to the distributed sources which
increases the fraction outside groups and clusters.

The objects returned to the stellar sample will also affect the Class
I/Class II ratio. Applying the density filter returns a sample with
$\sim$16\% Class I objects and the remainder Class II. If we apply the
combined filter, typically 4\% are Class I, 96\% Class II. The effect
on the ratio test is small: we increase the proportion of Class I
objects in the H\,{\sc ii} cavity from 4\% to $\sim$7\%. The fraction
associated with the molecular gas remains at 23\%, thus our earlier
result and conclusion still holds. A summary of the results of these
results is given in Table 7 below.

\begin{deluxetable}{lcccccc}
\tablecolumns{7}
\tablenum{7}
\tablewidth{0pt} 
\tabletypesize{\scriptsize}
\tablecaption{Clusters/Groups in W5}
\tablehead{ \colhead{} & \multicolumn{3}{c}{MST straight-line fit} & \multicolumn{3}{c}{MST $N_{grp}$---$d_s$ maximum} \\
\colhead{Filter} & \colhead{\% Clustered} & \colhead{$d_s$ (pc)} & \colhead{$N_{grp}$} & \colhead{\% Clustered} & \colhead{$d_s$ (pc)} & \colhead{$N_{grp}$}}
\startdata
Space Density & 66.0 & 0.8 & 19 & 43.4 & 0.5 & 26 \\
Combined & 68.0 & 0.8 & 17 & 45.5 & 0.5 & 25 \\
No AGN & 70.0 & 0.86 & 16 & 44.2 & 0.54 & 27
\enddata
\end{deluxetable}

For the smaller sample of PAH galaxies, statistical analysis carried
out as above predicted on average 211 objects. However, we find 198
candidate PAH galaxies in W5, of which certainly some fraction are
stellar objects as can be seen from the non-uniformity in Figure
13. Although our model predicts an excess of PAH galaxies over that
detected, we can still use the same procedure for these objects as for
the AGN to generate a probability P(YSO), based on colors and space
density. We present the 198 PAH galaxy candidates in Table 6 along
with the AGN with the same format of coordinates and photometry. We
provide a P(YSO) value for each, derived from local space density and
$[3.6] - [5.8]$ and $[4.5] - [8.0]$ colors.

%\begin{figure}
%\begin{center}
%\includegraphics[width=5in]{pah_nnd6.eps}
%\caption{Nearest neighbor distance distribution for objects classified as PAH galaxies by our pipeline.}
%\end{center}
%\end{figure}

%\bibliographystyle{apj} \bibliography{koenig}

\end{document}